\documentclass[11pt,a4paper]{article}
\pdfoutput=1
\usepackage{jheppub}
\usepackage[export]{adjustbox}
\usepackage{amssymb,amsmath}
\usepackage{hyperref}
\usepackage{color}
\usepackage{graphicx}
\usepackage{ulem}
\usepackage{graphics, color}
\usepackage{overpic}
\usepackage{float}
\usepackage{hyperref}
\usepackage[english]{babel}
\usepackage{hyperref}
\usepackage{amsfonts}
\usepackage[usenames,dvipsnames]{xcolor}
\usepackage{amsmath,amsfonts,amssymb,verbatim,float,mathtools}
\usepackage{physics}
\usepackage{tikz}

\def\XXint#1#2#3{{\setbox0=\hbox{$#1{#2#3}{\int}$}
\vcenter{\hbox{$#2#3$}}\kern-.5\wd0}}

\begin{document}
\title{Coherence effects in disordered geometries with a field-theory dual}
\author[a,b]{Tom\'as Andrade}
\emailAdd{tomas.andrade@physics.ox.ac.uk}
\affiliation[a]{Rudolf Peierls Centre for Theoretical Physics, University of Oxford, 1 Keble Road, Oxford OX1 3NP, UK}
\affiliation[b]{Departament de F\'isica Qu\`antica i Astrof\'isica, Institut de Ci\`encies del Cosmos (ICCUB), Universitat de Barcelona, 
Mart\'i i Franqu\`es 1, E-08028 Barcelona, Spain}
\author[c]{Antonio M. Garc\'\i a-Garc\'\i a}
\emailAdd{amgg@sjtu.edu.cn}
\affiliation[c]{Shanghai Center for Complex Physics, 
	Department of Physics and Astronomy, Shanghai Jiao Tong
	University, Shanghai 200240, China}

\author[d]{Bruno Loureiro}
\emailAdd{bl360@cam.ac.uk}
\affiliation[d]{TCM Group, Cavendish Laboratory, University of Cambridge, JJ Thomson Avenue, Cambridge, CB3 0HE, UK}

\abstract{
We investigate the holographic dual of a probe scalar in an asymptotically Anti-de-Sitter (AdS) disordered background which is an exact solution of Einstein's equations in three bulk dimensions. Unlike other approaches to model disorder in holography, we are able to explore quantum wave-like interference effects between an oscillating or random source and the geometry. In the weak-disorder limit, we compute analytically and numerically the one-point correlation function of the dual field theory for different choices of sources and backgrounds. The most interesting feature is the suppression of the one-point function in the presence of an oscillating source and weak random background. We have also computed analytically and numerically the two-point function in the weak disorder limit. We have found that, in general, the perturbative contribution induces an additional power-law decay whose exponent depends on the distribution of disorder. For certain choices of the gravity background, this contribution becomes dominant for large separations which indicates breaking of perturbation theory and the possible existence of a phase transition induced by disorder.}

\maketitle

%%%%%%%%%%%%%%%%%%%%%%%%%%%%%%%%%%%%%%%%
\section{Introduction}
%%%%%%%%%%%%%%%%%%%%%%%%%%%%%%%%%%%%%%%%
Holographic dualities, that relate classical theories of gravity to strongly-coupled quantum field theories, are now a forefront research area not only in high energy physics but also in quantum information and condensed matter physics. 
In the latter, it is emerging as a powerful tool to describe universal properties of strongly-correlated quantum systems. One of the main challenges for the application of holographic techniques in this context is the description of disorder, which is ubiquitous in realistic systems and directly responsible for a broad variety of phenomena ranging from momentum relaxation to quantum interference leading to different forms of localization \cite{Anderson1958,Abou-Chacra1973,Basko2006a,john1984,anderson1985}. The introduction of disorder in gravity backgrounds with a negative cosmological constant relevant for holography requires the solution of spatially inhomogeneous Einstein's equations, in general a difficult task.

Different approximation schemes have been proposed to make the problem technically tractable while keeping some of the expected phenomenology related to the introduction of disorder. 
For instance, momentum relaxation, a rather general consequence of any form of disorder, 
can be achieved by adding a massless scalar \cite{Bardoux2012, Andrade2014, Andrade2015, Andrade2016} that depends linearly on the boundary coordinates. Since the scalar field only couples to gravity through its derivatives, translation invariance is broken by the background but the equations of motion are still independent of the spatial coordinates which facilitates substantially the calculation, in many cases analytical, of transport properties.

 As was expected, the electrical conductivity is always finite and depends directly on the strength of the translational symmetry breaking characterized by the slope, with respect to the spatial coordinates, of the scalar field in the boundary. For weak momentum relaxation, the electrical conductivity reproduces the expected phenomenology of Drude's model,  which includes a peak at low frequencies, remnant of the broken translational symmetry, followed by a decay for higher frequencies. For stronger relaxation the Drude peak is suppressed, leading to an incoherent 'bad-metal' behaviour \cite{Kim2014, Davison2014, Amoretti2014, Davison2015a}. However even in the limit of infinite relaxation no insulating behaviour is observed. Recently, models that consider the coupling of the scalar field to the gravity and the Maxwell term managed to reproduce a vanishing conductivity in the limit of infinite relaxation \cite{Baggioli2016, Gouteraux2016, Garcia-Garcia2016a}. Yet, in this limit the effective charge in these models vanish, so strictly speaking they cannot be considered insulators. Other effective models of momentum relaxation in holography include the memory matrix formalism \cite{Hartnoll2008b, Davison2013a, Lucas2015a, Lucas2015}, helical and Q-lattices \cite{Donos2012,Donos2013, Donos2014} and massive gravity \cite{Vegh2013, Blake2013, Davison2013, Baggioli2014}. Similar results \cite{Aprile2014, Donos2014b, Kachru2010, Horowitz2012, Horowitz2012a, Ling2013, Chesler2013} have been obtained even for Maxwell fields with a spatially oscillating chemical potentials in the boundary leading to inhomogeneous Einstein's equations. We note that effects such as localization are precluded by design since a random but homogeneously distributed, and therefore delocalized, chemical potential is an input in this approach.  

Disorder has also been introduced at the level of the action by a random source coupled to the dual conformal field theory operator \cite{Fujita2008, Aharony2015}. By using the replica trick, it is possible to integrate out the random coupling resulting in a double trace deformation of the non-disordered theory. For marginal perturbations, renormalization group techniques suggest the existence of logarithmic corrections in the two-point correlation function that spoils conformal symmetry. Interestingly, this is in agreement with the expected behavior of certain two dimensional conformal field theories perturbed by disorder (see \cite{cardy2013} and references therein). 

%Explicit inhomogeneous fields have been studied before in the Holography literature. %Inhomogeneous holographic lattices have been considered analytically \cite{Chesler2013, %Aprile2014, Donos2014b} and numerically \cite{Kachru2010, Horowitz2012, Horowitz2012a, %Ling2013, Chesler2013, Aprile2014}, and they were shown to reproduce some features of the %aforementioned effective approaches \cite{Blake2014}. 
Another popular approach is to consider a random field in the boundary, in most cases a random chemical potential (source) for gauge (scalar) fields, but neglecting the backreaction in the gravity background (probe limit) \cite{Ryu2011, Saremi2012, Zeng2013, Taylor2014, Arean2014, Arean2014a,Arean2015,Lucas2015b,Araujo2016}. Analytical attempts to go beyond this probe limit have found logarithmic infra-red (IR) divergences in the gravity background, signaling the breakdown of perturbation theory \cite{Adams2011, Adams2014}. A resummation scheme for the divergent expansion was proposed in \cite{Hartnoll2014a,Hartnoll2015,Hartnoll2015a,Hartnoll2016} resulting in an averaged gravity background with an emerging Lifshitz scaling symmetry, which was shown to be a generic feature of marginal disordered deformations \cite{OKeeffe2015, Garcia-Garcia2016}. 
For geometries with a horizon, the most general result in this context is that of Gauntlett, Donos and collaborators  \cite{Donos2013a,Donos2014b,Donos2014c,Donos2015,Donos2015a,Donos2017} who found closed expressions for the dc conductivity and other averaged transport coefficients in generic inhomogeneous backgrounds. As a direct consequence, bounds on the electric and thermal conductivity were proposed \cite{Grozdanov2015, Grozdanov2015a,Ikeda2016,Lucas2015c}. 

The conclusion of this more direct approach is similar to the one from the phenomenological models of momentum relaxation discussed previously, namely, it is not possible to reach an insulating state which is believed to be a distinctive feature of strong disorder in condensed matter systems. Even simpler coherence effects such as weak-localization \cite{Altshuler1980}, which are precursors of a metal-insulator transition, have not yet been clearly identified. 

In this manuscript we propose a new approach to model disorder in holography which has the potential to reproduce some of these coherent effects. We switch perspectives and consider the effect of a disordered geometry with no horizon on a probe scalar field. This is accomplished by considering a family of three dimensional random geometries that solve Einstein's equations exactly\footnote{The fact that we can find an 
exact solution of Einstein's equations involving a free function is due to the fact that all vacuum solutions are pure diffeomorphisms in three 
dimensions. The generalization to higher dimensions would involve solving the equations numerically.} 
and neglecting the backreaction of the scalar in the geometry. This family is indexed by a parameter which we take to be a random function of the boundary coordinates. 
The scalar field feels the geometry as an effective inhomogeneous coefficient in the equations of motion. This is reminiscent, though we cannot establish a precise mapping, of a one-dimensional wave equation with a random refractive index \cite{anderson1985,john1984} plus additional terms that control the evolution in the radial direction which are related to interactions in the boundary. 
 
According to the holographic dictionary, the geometry is dual to a strongly-coupled disordered plasma living at the boundary, and the scalar field sources a boundary dual operator. We investigate numerically and analytically the properties of this disordered plasma by looking at one and two-point functions of this scalar operator. For the one-point function our main result is the observation of coherence effects, due to interference between an oscillating source and the random geometry, that, in some cases, leads to the strong suppression of oscillations even for weak disorder. The contribution to the two-point function for a weak random Gaussian geometry is still a power-law decay for large distances with an exponent that depends on both the scalar mass and disorder correlations modeled by a non-trivial power spectrum. In some cases, this correction becomes dominant for large distances which suggests the breaking of perturbation theory and the possible transit of the system to a new disorder-driven fixed point.

The manuscript is organized as follows. In the next section we introduce the geometry we will be studying and discuss the equation of motion for the probe scalar. In section \ref{sec:perturb} we solve these equations analytically in the limit of a weak-disordered background. The one and two-point functions of the dual operator are computed for different choices of both sources and inhomogeneous geometries. In section \ref{sec:nume}, we solve the equation of motion by numerical techniques and compute in certain cases the one and two-point correlation function of dual scalar. Section \ref{sec:comp} is devoted to a comparison of our results with previous approaches to disorder in holography. We conclude in section \ref{sec:discu} with a summary of results and ideas for future work. The appendices offer a wider discussion of technical points which are used throughout the main body of the paper.

%%%%%%%%%%%%%%%%%%%%%%%%%%%%%%%%%%%%%%%%
\section{Setup}
%%%%%%%%%%%%%%%%%%%%%%%%%%%%%%%%%%%%%%%%
\label{sec:setup}
In this section we introduce our objects of study. First, we introduce a family of geometries in $d+1=3$ spacetime dimensions. This family is characterised by an arbitrary function which we can take to be random. We next introduce a minimally coupled scalar field and discuss its equation of motion and associated boundary conditions. The aim is to use this field to probe the properties of the inhomogeneous geometry, which holographically can be interpreted as a strongly-coupled disordered field theory.

\subsection{Geometry}

Solutions of the vacuum $d+1$-dimensional Einstein's Equations with a negative cosmological constant have been classified in the pioneering work of Fefferham and Graham \cite{Fefferman2007, Fefferman2007b}. For $d>2$, we can integrate Einstein's Equations in a neighborhood of the boundary by requiring that the Weyl tensor vanish. This condition constrains the boundary metric to be conformally flat. However, in $d=2$ the Weyl tensor vanishes exactly, leaving the conformal class of the boundary metric arbitrary \cite{Skenderis2000}. 

In this manuscript we will be mainly interested in the following family of metrics defined by a global coordinate patch $x^a = (\rho, t, x)$ as
\begin{align}
\label{eq:bulk_metric}
\dd s^2 = \frac{\dd \rho^2}{4\rho^2}+\frac{1}{\rho}\left(-\dd t^2 + \dd x^2 + 2 g_{tx}(x)\dd t \dd x\right),
\end{align}
\noindent where $g_{tx}(x)$ is an arbitrary function of the boundary coordinate $x$. In line with our discussion above, it is easy to check that this family satisfies Einstein's Equations in $d=2$ dimension with a negative cosmological constant for any $g_{tx}$. In these coordinates, the conformal boundary is located at $\rho=0$ and the induced conformal metric $g^{(0)}$ is given by $\rho \dd s^2|_{\rho = 0} = -\dd t^2 + \dd x^2 + 2g_{tx}(x)\dd t \dd x$. The Poincare horizon is parametrized by $\rho = \infty$. 

From a holographic perspective, this space-time encodes the degrees of freedom of a strongly-coupled field theory living on the boundary metric described above. However, the dual stress tensor vanishes since this change in the boundary metric does not induce sub-leading terms in the bulk metric. This resembles the situation we encounter in the axion model of \cite{Andrade2014}, where the marginal, spatially dependent, scalar sources do not excite a vev.
In the next sections we will be interested in studying the family of geometries in Eq.\eqref{eq:bulk_metric} for different choices of $g_{tx}$. In particular, we will be interested in the case where $g_{tx}(x)$ is a random Gaussian process, as introduced in Appendix \ref{app:A}. We will study the dynamics of a minimally coupled scalar field as a way of probing the effects of the disordered geometry. The aim is to get an insight into the nature of the strongly-coupled disordered dual field theory.

\subsection{Scalar Field and Equations of Motion}
We consider a probe scalar field $\psi$ of mass $m$ minimally coupled to the geometry in Eq.\eqref{eq:bulk_metric}. The equation of motion is given by $(\Delta_{g}-m^2)\psi=0$, where the curved Laplacian can be written in a chart $x^a$ as $\Delta_g = \frac{1}{\sqrt{-g}}\partial_a\left(\sqrt{-g}~g^{ab}\partial_b\right)$. Since $\partial_t$ is a killing vector for Eq.\eqref{eq:bulk_metric}, we can restrict our attention to static configurations $\psi = \psi(\rho, x)$. The equation of motion thus reads
\begin{align}
\label{eq:scalar_eom}
4\rho^2 \partial_{\rho}^2\psi -\rho\frac{g_{tx}\partial_x g_{tx}}{(1+g_{tx}^2)^2}\partial_{x}\psi + \rho\frac{1}{1+g_{tx}^2}\partial_{x}^2\psi -m^2\psi=0.
\end{align}
We are interested in solutions satisfying the following boundary conditions,
\begin{align}
	\lim\limits_{\rho\to\infty}\psi(\rho,x)& <\infty,\label{bc1}\\
	\lim\limits_{\rho\to 0}\rho^{\frac{\nu-1}{2}}\psi(\rho,x)&=s(x) \label{bc2},
\end{align}
\noindent where we have defined $\nu = \sqrt{1+m^2}$. The first boundary condition assures regularity of the solution at the Poincar\'e horizon $\rho=\infty$. As discussed in Appendix \ref{app:B}, the second boundary condition defines a field $s(x)$ living in the boundary $\rho = 0$ with conformal dimension $\Delta_{-} = 1-\nu$. This field sources a dual boundary operator $\mathcal{O}(x) \sim \rho^{-\frac{\nu+1}{2}}\psi$ of conformal dimension $\Delta_{+}=1+\nu$.

Even for simple choices of $g_{tx}$, Eq.\eqref{eq:scalar_eom} remains largely intractable analytically. However when $g_{tx}$ is small we can compute perturbative corrections to the plain AdS$_3$ result analytically, helping us to build an intuition of the effects of the weakly disordered geometry. We will recur to numerical methods to test the analytical prediction and also to explore the region of stronger disorder not accessible to an analytical treatment.

%%%%%%%%%%%%%%%%%%%%%%%%%%%%%%%%%%%%%%%%
\section{Perturbative analytical calculation of one-point and two-point scalar correlation functions}
%%%%%%%%%%%%%%%%%%%%%%%%%%%%%%%%%%%%%%%%
\label{sec:perturb}
In this section we study Eq.\eqref{eq:scalar_eom} perturbatively for different choices of $g_{tx}$. For convenience, we will set $m^2 = -\frac{3}{4}$ throughout this section, which is equivalent to choosing $\nu = \frac{1}{2}$. Note this choice respects the Breitenlohner-Freedman bound, 
$m^2 \geq  -1$ in $d= 2$, \cite{Breitenlohner:1982bm, Breitenlohner:1982jf}.
We will divide our discussion by order in perturbation theory, and focus in two observables: the expectation value of the dual boundary operator (one-point function) and the two-point function. We can obtain both these quantities by solving the perturbative equations with the appropriate boundary conditions.

\subsection{Zeroth Order}
To set up the perturbative analysis, let $g_{tx}(x) = \epsilon~ g(x)$ for a parameter $\epsilon\ll 1$ measuring the amplitude of $g_{tx}$ fluctuations. For $\epsilon=0$, Eq.\eqref{eq:scalar_eom} reduces to the equation of a massive scalar field in AdS$_3$ given by
\begin{align}
4\rho^2\partial_{\rho}^2\psi_{(0)}+\rho\partial_x^2\psi_{(0)} +\frac{3}{4} \psi_{(0)} = 0.	
\end{align}
Letting $\psi_{(0)}(\rho,x) = \int_{\mathbb{R}}\frac{\dd k}{2\pi} e^{ikx}f_{k}(\rho)$, the equation above reduces to
\begin{align}
\label{eq:0eq}
4\rho^2f''_{k}-\left(\rho k^2-\frac{3}{4}\right) f_k = 0,
\end{align}
\noindent which has general solution
\begin{align}
\label{eq:0thhom}
f_{k}(\rho,x)= a_{k}\rho^{1/4} e^{-k\sqrt{\rho}}+b_{k} \rho^{1/4} e^{k\sqrt{\rho}}. 
\end{align}
Regularity at the Poincar\'e horizon Eq.\eqref{bc1} requires that $a_k=0$ for $k<0$ and $b_k = 0$ for $k>0$, which can be written compactly as $f_{k}(\rho) = a_k \rho^{1/4} e^{-|k|\sqrt{\rho}}$ for a different constant $a_k$. Now letting $s(x) = \int_{\mathbb{R}}\frac{\dd k}{2\pi} e^{ikx} s_{k}$, boundary condition Eq.\eqref{bc2} can be imposed coefficient wise to yield $a_k = s_k$. The full solution therefore reads
\begin{align}
\label{eq:0thsol}
\psi_{(0)}(\rho, x) =\rho^{1/4} \int_{\mathbb{R}}\frac{\dd k}{2\pi} e^{ikx}e^{-|k|\sqrt{\rho}} s_k.
\end{align}
Note that this depends directly on the Fourier components of the source. We are interested in a few particular cases that we outline below.

\begin{description}
	\item[Constant source] If $s(x) = s = \int_{\mathbb{R}}\frac{\dd k}{2\pi} e^{ikx}2\pi\delta(k)s$ is constant, we have $\psi_{(0)}(\rho, x) = \rho^{1/4} s$. This sources a dual boundary operator with expectation value $\langle\mathcal{O}(x) \rangle = 2\nu\lim\limits_{\rho\to 0}\rho^{-\frac{\nu+1}{2}}\psi_{(0)} = 0$.
	\item[Oscillating source] If $s(x) = s\cos{qx} = s\int_{\mathbb{R}}\frac{\dd k}{2\pi}e^{ikx}\pi\left[\delta(k-q)+\delta(k+q)\right]$ for $q\in\mathbb{R}$, we have $\psi_{(0)}(\rho, x) = s\rho^{1/4}e^{-|q|\sqrt{\rho}}\cos{qx}$. This sources a dual boundary operator with expectation value given by $\langle \mathcal{O}(x)\rangle = -s|q| \cos{qx}$.
	\item[Superposition of oscillations] Consider now a source given by a superposition of $N$ oscillating modes $s(x) = \sum\limits_{n=1}^{N} s_{n} \cos(q_n x+\gamma_n)$, where $s_n$, $q_n$ and $\gamma_n$ can be freely chosen. Noting that the finite sum can be exchanged with the integral and following the same steps as above mode-wise we find $\psi_{(0)}(\rho,x) = \rho^{1/4}\sum\limits_{n=1}^{N} s_n e^{-|q_n|\sqrt{\rho}}\cos(q_n x+\gamma_n)$. The one-point function thus reads $\langle \mathcal{O}(x)\rangle = -\sum\limits_{n=1}^{N}s_n |q_n| \cos(q_n x+\gamma_n)$.
	\item[Delta source and two-point function] As discussed in Appendix \ref{app:C}, the boundary-to-bulk propagator is given by solving the equations of motion with a delta source, $s(x) = \delta(x) = \int_{\mathbb{R}}\frac{\dd k}{2\pi} e^{ikx}$. We thus have,
	\begin{align}
	K_{(0)}(\rho, x-y) = \rho^{1/4}\int_{\mathbb{R}}\frac{\dd k}{2\pi} e^{ik(x-y)} e^{-|k|\sqrt{\rho}} = \frac{1}{\pi} \frac{\rho^{3/4}}{(x-y)^2+\rho}.
	\end{align}
\noindent which follows the shape of a Lorentzian (or Cauchy) distribution. Following the discussion in Appendix \ref{app:B}, the boundary two-point function can be obtained by
\begin{align}
\label{order0static}
\langle\mathcal{O}(x)\mathcal{O}(y)\rangle=2\nu\lim\limits_{\rho\to 0}\rho^{-\frac{1+\nu}{2}}K_{(0)}(\rho;x-y) = \frac{1}{\pi} \frac{1}{(x-y)^2}. 
\end{align}
Note this is not the expected result for an operator of conformal dimension $2\Delta_{+} = 3$ but rather for an operator of conformal dimension $2\Delta_{+} =2$. This is because, by considering a static field, we effectively reduce the conformal dimension of the problem. Since we will be interested in static inhomogeneous configurations in what follows, this is the object we will be computing corrections for.
\end{description}

\subsection{Second Order}
Note that since $g_{tx}$ appears only quadratically in Eq.\eqref{eq:scalar_eom}, there are no non-trivial order one corrections to $\psi_{(0)}$. It is thus sufficient to consider $\psi = \psi_{(0)}+\epsilon^2 \psi_{(2)}+O(\epsilon^4)$. Inserting into Eq.\eqref{eq:scalar_eom} and expanding up to second order leads to
\begin{align}
\label{eq:2ndorder}
	4\rho^2\partial_{\rho}^2\psi_{(2)} + \rho~\partial_{x}^2\psi_{(2)} +\frac{3}{4}\psi_{(2)}=\rho ~g(x)^2\partial_{x}^2\psi_{(0)}+\rho~ g(x)\partial_x g(x)\partial_{x}\psi_{(0)}.
\end{align}
Note that the zeroth order solution act as a source for the perturbative correction. Since the full solution has to satisfy the boundary conditions, we have to apply them order by order. For instance every term in the $\epsilon$-expansion has to be regular at the Poincar\'e horizon. However since we have enforced boundary condition Eq.\eqref{bc2} at zeroth order, we have to set $\lim\limits_{\rho\to 0}\rho^{\frac{\nu-1}{2}}\psi_{(2)}=0$ for the inhomogeneous geometry not to correct the fixed boundary source $s(x)$. Summarizing, we have to solve Eq.\eqref{eq:2ndorder} subjected to 
\begin{align}
	\lim\limits_{\rho\to\infty}\psi_{(2)}(\rho,x)& <\infty,\label{2ndbc1}\\
	\lim\limits_{\rho\to 0}\rho^{\frac{\nu-1}{2}}\psi_{(2)}(\rho,x)&=0 \label{2ndbc2}.
\end{align}
If $g_{tx}$ is of Schwarz class, we can attempt to solve Eq.\eqref{eq:2ndorder} in Fourier space as we did for the zeroth-order result. Letting $\psi_{(2)}(\rho, x) = \int_{\mathbb{R}}\frac{\dd k}{2\pi} e^{ikx} f_{k}(\rho)$ and $g(x) = \int_{\mathbb{R}}\frac{\dd k}{2\pi}e^{ikx}g_k$, we can rewrite Eq.\eqref{eq:2ndorder} in Fourier space as
\begin{align}
\label{eq:2ndFT}
4\rho^2 f_{k}''-\left(\rho k^2-\frac{3}{4}\right)f_k &=-\frac{\rho^{5/4}}{2}\int_{\mathbb{R}}\dd l\int_{\mathbb{R}}\dd q~(k-l)(2k-l) e^{-|k-l|\sqrt{\rho}}g_{l-q}g_q s_{k-l}
\end{align}
\noindent where the right-hand side has been evaluated applying the convolution theorem with the zeroth order solution Eq.\eqref{eq:0thsol}\footnote{Note that the convolution representation is not unique, but all representations are equivalent up to a translation in the momentum integration.}. This integral-differential equation cannot be solved in a closed form. In the following subsections we discuss solutions for specific inhomogeneous configurations. For each choice of $g$, we can consider different choices of source and study the interplay between the source, the geometry and the resulting expectation value and two-point function of the dual boundary operator.

\subsubsection{Constant geometry}
\label{sec:constant}
The simplest example is given by taking $g(x) = 2\pi~g$ to be a constant. In this case $g_{k} = g\delta(k)$ and for generic source
\begin{align}
4\rho^2 f_{k}''-\left(\rho k^2-\frac{3}{4}\right)f_k &=-\rho^{5/4}k^2 e^{-|k|\sqrt{\rho}}s_{k}g^2.
\end{align}
Note that the linear differential operator on the left-hand side is exactly the same as for the zeroth-order equation. This is generic and hold at all orders in perturbation theory. The only difference is the source term on the right-hand side. By linearity, the general solution will be a linear combination of the solution for the homogeneous equation plus a particular solution. As before, the homogeneous solution has one exponentially diverging piece which should be set to zero by regularity at the Poincar\'e horizon. This leads to
\begin{align}
f_{k}(\rho) = a_k \rho^{1/4}e^{-|k|\sqrt{\rho}}+\frac{g^2}{4}\rho^{1/4} e^{-|k|\sqrt{\rho}} \left(1+2|k|\sqrt{\rho}\right)s_k,
\end{align}
\noindent where $a_k$ is an integration constant. Close to the boundary $\rho=0$, the solution behaves as
\begin{align}
f_{k}(\rho) \underset{\rho= 0}{\sim} \left(a_k+\frac{g^2}{4}s_k\right)\rho^{1/4}+O\left(\rho^{3/4}\right),	
\end{align}
\noindent and therefore boundary condition Eq.\eqref{2ndbc2} imposes $a_k = -\frac{g^2}{4}s_k$. Finally, we can write
\begin{align}
\psi_{(2)}(\rho,x)=\frac{1}{2}\rho^{3/4}g^2\int_{\mathbb{R}}\frac{\dd k}{2\pi}e^{ikx} |k|e^{-|k|\sqrt{\rho}} s_k.
\end{align}
As in the zeroth-order solution, we discuss below a few cases of interest.
\begin{description}
	\item[Constant source] Let $s(x) = s$ be a constant. Inserting in the above leads trivially to $\psi_{(2)}(\rho,x)=0$. Therefore the off-diagonal constant metric does not affect the zeroth order boundary one-point function.
	\item[Oscillating source] Let $s(x) = s\cos{qx}$. 
%	
%	Inserting in the above leads to $\psi_{(2)}(\rho,x) = \frac{\rho^{3/4}g^2}{2}|q|e^{-|q|\sqrt{\rho}}s\cos{qx}$. 
	Inserting in the above leads to 
\begin{equation*}
  \psi_{(2)}(\rho,x) = \frac{\rho^{3/4}g^2}{2}|q|e^{-|q|\sqrt{\rho}}s\cos{qx}. 
\end{equation*}	
	At the boundary $\rho=0$ this induces a correction to the zeroth-order one-point function which is proportional to $g$,
	\begin{align}
	\langle\mathcal{O}(x)\rangle = -|q|\left(1-\frac{\epsilon^2}{2}g^2+O(\epsilon^4)\right) s\cos{qx}.
	\end{align}
	\item[Superposition of oscillations] Let $s(x)=\sum\limits_{n=1}^{N}s_k \cos\left(q_n x+\gamma_n\right)$. This case is similar to the above, since we can integrate term by term in the sum to give, 
\begin{equation*}
	\psi_{(2)}(\rho, x) = \rho^{3/4}\frac{g^2}{2}\sum\limits_{n=1}^{N}|q_n|e^{-|q_n|\sqrt{\rho}}s_n \cos(q_n x+\gamma_n).
\end{equation*}	
 The correction to the boundary one-point reads
	\begin{align}
	\langle\mathcal{O}(x)\rangle = -\sum\limits_{n=1}^{N}|q_n|\left(1-\frac{\epsilon^2}{2}g^2+O(\epsilon^4)\right) s_n\cos(q_n x+\gamma_n).
	\end{align}
	Again, it represents just a renormalization of the amplitude of the zeroth one-point function.
	\item[Delta source and two-point function] Recall that, as discussed in the previous section and in the Appendix \ref{app:C} that the boundary-to-bulk propagator $K(\rho; x, y)$ can be computed by solving the equations of motion with $s(x)=\delta(x)$. We thus have $K_{(2)}(\rho; x) = \frac{g^2}{2\pi}\rho^{3/4}\frac{x^2-\rho}{(\rho+x^2)^2}$. Evaluating at the boundary leads to a correction to the first zeroth-order two-point function,
	\begin{align}
	\label{gconstcorr}
	\langle\mathcal{O}(x)\mathcal{O}(y)\rangle = \left(1-\frac{\epsilon^2}{2}g^2+O(\epsilon^4)\right)\frac{1}{\pi}\frac{1}{|x-y|^2}.
	\end{align}	
	\end{description}

Note that for all the cases above the constant off-diagonal metric element perturbatively decreases the amplitude of the boundary one-point function and two-point function. Since in perturbation theory the equations are linear this case can be interpreted as the mean result for a inhomogeneous geometry. The amplitude damping raises the question on whether in a non-perturbative setup the geometry can effectively suppress the boundary correlation functions.

\subsubsection{Oscillating geometries}
\label{sec:oscillating}
We now consider a generic superposition of $N$ oscillating modes, $g(x) = \sum\limits_{n=1}^{N}A_{n}e^{i \omega_n x}$ for $\omega_n \in\mathbb{R}$. For example, an interesting particular case is $\omega_1 = -\omega_2 = \omega$, $A_1 = A_2 = \frac{g}{2}$ and $A_n = 0$ for $n>2$ which correspond to $g(x) = g\cos{\omega x}$. The Fourier modes are given by a Dirac comb $g_k = \sum\limits_{n=1}^{N}A_n\delta(k-\omega_n)$ and for a generic source the equations of motion in Fourier space read,
\begin{align}
4\rho^2 f_{k}''-\left(\rho k^2-\frac{3}{4}\right)f_k =-\rho^{5/4}\sum\limits_{n=1}^{N}\sum\limits_{m=1}^{N} A_n A_m (k-&\omega_n-\omega_m)(2k-\omega_n-\omega_m)\times\notag\\&\times e^{-|k-\omega_n-\omega_m|\sqrt{\rho}}s_{k-\omega_n-\omega_n}.
\end{align}
By linearity, we can solve the equation above term by term. The regular solution at the Poincar\'e horizon is given by the homogeneous solution Eq.\eqref{eq:0thhom} plus the sum of each individual particular solution
\begin{align}
f_{k}(\rho)=\frac{\rho^{1/4}}{2}\sum_{n=1}^{N}\sum\limits_{m=1}^{N}A_n A_m \frac{k-\omega_n-\omega_m}{\omega_n+\omega_m}\left[e^{-|k-\omega_n-\omega_m|\sqrt{\rho}}-e^{-|k|\sqrt{\rho}}\right]s_{k-\omega_n-\omega_m}.
\end{align}
Taking the Fourier transform, we arrive at an implicit solution for the generic source
\begin{align}
\label{eq:2ndosci}
\psi_{(2)}(\rho, x)	= \frac{\rho^{1/4}}{2}\sum_{n=1}^{N}\sum\limits_{m=1}^{N}\frac{A_n A_m}{\omega_n+\omega_m} \int_{\mathbb{R}}\frac{\dd k}{2\pi}e^{ikx}(k-\omega_n-\omega_m)\left[e^{-|k-\omega_n-\omega_m|\sqrt{\rho}}-\right. &\left. e^{-|k|\sqrt{\rho}}\right]\times\notag\\\times s_{k-\omega_n-\omega_m}.
\end{align}
As before, we now analyse some interesting particular cases where the above integral can be done explicitly.
\begin{description}
	\item[Constant source] Let $s(x)=s$ be a constant, i.e. $s_k = \delta(k)$. As before, the above integral trivially gives $\psi_{(2)}=0$. There are no corrections to the boundary dual operator.
	
	\item[Plane wave source] Let $s(x)=s e^{iqx}$, i.e. $s_k = s\delta(k-q)$. The Fourier transform is given by
	\begin{align}
	\psi_{(2)}(\rho, x) = \rho^{1/4}s e^{iqx}\frac{q}{2}\sum_{n=1}^{N}\sum\limits_{m=1}^{N}\frac{A_n A_m}{\omega_n+\omega_m} e^{i(\omega_n+\omega_m)x}\left[e^{-|q|\sqrt{\rho}}-e^{-|\omega_n+\omega_m+q|\sqrt{\rho}}\right].
	\end{align}
 	Close to the boundary, this gives a correction to the dual one-point function
 	\begin{align}
 	\delta\langle\mathcal{O}(x)\rangle = s e^{iqx}\frac{q}{2}\sum_{n=1}^{N}\sum\limits_{m=1}^{N}\frac{A_n A_m}{\omega_n+\omega_m} e^{i(\omega_n+\omega_m)x}\left(|\omega_n+\omega_m+q|-|q|\right).
 	\end{align}
	There are a couple of particular cases of special interest, 
	\begin{itemize}
	\item  $g(x) = g e^{-i\omega x}$ ($A_1 = g$, $\omega_1 = -\omega$, all other zero):
	\begin{align}
	\langle\mathcal{O}(x)\rangle &= -q\left(1-\frac{\epsilon^2}{4\omega}  \left(|q-2\omega|-|q|\right)g^2 e^{-2i\omega x}\right)s e^{iqx},\notag\\
	&=-q\left(1-\frac{\epsilon^2}{4\omega}  \left(|q-2\omega|-|q|\right)g(x)^2\right)s(x).
	\end{align}
	Consider $q,\omega\geq 0$ and define the relative correction $\Delta(x) = \frac{\delta \langle\mathcal{O}\rangle}{\langle\mathcal{O}\rangle_{(0)}}$, where $\langle\mathcal{O}\rangle_{(0)}$ is the zeroth order one-point result.
	We can identify three distinct regimes. For $q>2\omega$, the relative correction to the one-point function is positive and simply proportional to the geometry $\Delta(x)_{q\geq 2\omega}=-\frac{\epsilon^2}{2} g(x)^2 =-\frac{1}{2}g_{tx}(x)^2$, while for $q<2\omega$ the result depends explicitly on the relative strength of the modes, $\Delta(x)_{q<2\omega} = -\frac{\epsilon^2}{2}\left(\frac{q}{\omega}-1\right)g(x)^2 =-\frac{1}{2}(\frac{q}{\omega}-1)g_{tx}^2$. For $q<\omega$ the relative correction will be positive, while in the window $\omega<q<2\omega$ it becomes negative. Therefore the inhomogeneous geometry can either suppress ($q>\omega$) or enhance ($q<\omega$) the expectation value of the dual operator, depending on the coherence between the modes. The discussion is summarized in Fig.\ref{fig:1}.
\begin{figure}[H]
\centering
\begin{tikzpicture}
\draw [thick, ->](0,0) -- (6,0);
\foreach \x/\xtext in {0/0,2/$\omega$, 4/$2\omega$}{
\draw (\x cm,-2pt) -- (\x cm,2pt) node[above] {\xtext};}
\node[] at (1, +0.5)   (a) {\textbf{\textcolor{red}{$+$}}};
\node[] at (3, +0.5)   (a) {\textbf{\textcolor{red}{$-$}}};
\node[] at (5, +0.5)   (a) {\textbf{\textcolor{red}{$-$}}};
\end{tikzpicture}
\caption{Sign of the relative correction $\Delta(x)$ for different $q\geq 0$ and fixed $\omega$.}
\label{fig:1}
\end{figure}
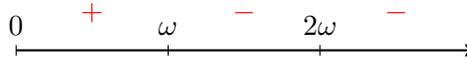
	
\item $g(x) = g \cos{\omega x}$ ($A_1 = A_2 = \frac{1}{2} g$, $\omega_1 = -\omega_2 = \omega$, all other zero):
	
	By linearity, this case can be obtained by summing the above with the reversed $q\to -q$. The relative correction $\Delta(x) = \frac{\delta \langle\mathcal{O}\rangle}{\langle\mathcal{O}\rangle_{(0)}}$ is given by
\begin{align}
\Delta(x) = -\epsilon^2\frac{g^2}{8} \left[\frac{|q+2\omega|-|q|}{2\omega}e^{2i\omega x}-\frac{|q-2\omega|-|q|}{2\omega}e^{-2i\omega x}+2\right].
	\end{align}
	Note that, different from the above in general it is not possible to rewrite the relative correction as an explicit function of the geometry. First, lets consider the case $q,\omega \geq 0$. Then $q+2\omega \geq 0$ and we have two subcases: $q\geq 2\omega$ and $q<2\omega$. These are given by
	\begin{align}
	\Delta(x)_{q\geq 2\omega} &= -\frac{\epsilon^2}{4} g^2\left(\cos{2\omega x}+1\right) = -\frac{\epsilon^2}{4} g^2\cos^2{\omega x} = -\frac{1}{2} g_{tx}^2\\
	\Delta(x)_{q<2\omega} &= -\frac{\epsilon^2}{8} g^2\left[e^{2i\omega x}+\left(\frac{q}{\omega}-1\right)e^{-2i\omega x}+2\right]
	\end{align}
	The two modes $e^{2i\omega}$ and $e^{-2i\omega}$ can interfere constructively or destructively depending on the relative sign of $q$ and $\omega$. Note that the correction is symmetric under $\omega \to -\omega$ as expected. The $q,\omega < 0$ case is similar up to a change of sign.
\end{itemize} 
		
	\item[Oscillating Source] We now build on the previous results to analyze more intricate cases. Let $s(x)=s \cos{qx}$. Linearity of Eq.\eqref{eq:2ndosci} together with the example above can be used to get the following correction,
	\begin{align}
 	\delta\langle\mathcal{O}(x)\rangle = \epsilon^2 s\frac{q}{4}\sum_{n=1}^{N}\sum\limits_{m=1}^{N}\frac{A_n A_m}{\omega_n+\omega_m}\Big[& \left. e^{i(\omega_n+\omega_m+q)x}\left(|\omega_n+\omega_m+q|-|q|\right)+\right.\notag\\ &\left. -e^{i(\omega_n+\omega_m-q)x}\left(|\omega_n+\omega_m-q|-|q|\right)\right].
	\end{align}
	Consider the particular subcases:
	\begin{itemize}
		\item $g(x) = g \cos{\omega x}$ ($A_1 = A_2 = \frac{1}{2} g$, $\omega_1 = -\omega_2 = \omega$, all other zero). The correction can be conveniently rearranged to give
		\begin{align}
		\delta\langle\mathcal{O}(x)\rangle = \epsilon^2q\frac{g^2 s}{4}\bigg[&\left.\frac{|q+2\omega|-|q|}{2\omega}\cos\left((2\omega+q)x\right)\right.\notag\\&\left.-\frac{|q-2\omega|-|q|}{2\omega}\cos\left((q-2\omega)x\right)+\cos{qx}\right],
		\end{align}
		\noindent which is, as expected, a real result. However note that the source modes are now coupled with the geometry. We can still write the relative correction $\Delta(x)$ by dividing by the source, and being careful to take into account that when the source vanishes the result is zero and not divergent. 
		\begin{align}
		\Delta(x) = -\epsilon^2\frac{g^2}{4}\left[\frac{|q+2\omega|-|q|}{2\omega}\frac{\cos\left((2\omega+q)x\right)}{\cos{q x}}-\frac{|q-2\omega|-|q|}{2\omega}\frac{\cos\left((q-2\omega)x\right)}{\cos{q x}}+1\right].
		\end{align}		
		This can be analyzed as before, giving
		\begin{align}
		\label{deltaoscx}
		\Delta(x)_{q\geq 2\omega} &= -\epsilon^2\frac{g^2}{4}\left[\frac{\cos\left((2\omega+q)x\right)}{\cos{q x}}+\frac{\cos\left((q-2\omega)x\right)}{\cos{q x}}+2\right]\notag\\ 
		 &= \epsilon^2\frac{g^2}{4}\left[\cos{2\omega x}+1\right]
		 = \frac{1}{2}g_{tx}^2,
		\end{align}
		\begin{align}
		\Delta(x)_{q< 2\omega} &=-\epsilon^2\frac{g^2}{4}\left[\frac{\cos{\left((2\omega+q)x\right)}}{\cos{qx}}+\left(\frac{q}{\omega}-1\right)\frac{\cos{\left((2\omega-q)x\right)}}{\cos{qx}}+1\right].
		\end{align}	
		Interestingly, as with all the examples we considered above, the relative correction for the case $q>2\omega$ factors into a contribution that only depends on the underlying geometry. This can be interpreted intuitively as follows. If the wavelength of the source ($\propto q^{-1}$) is much smaller than the wavelength of the geometry, the dual operator will not 'feel' the inhomogeneity, leading to a coupling similar to the constant geometry case discussed in Section \ref{sec:constant}. In Fig.\ref{fig:3} we plot the correction for different configurations of $(\omega, q)$ and inhomogeneity strength $\epsilon$.
	\begin{figure}[htp]
	\centering
	\includegraphics[width=.5\textwidth]{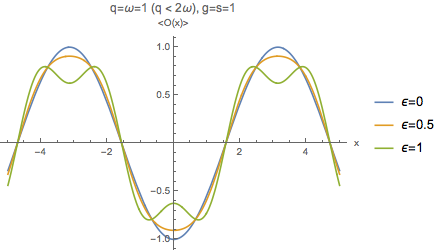}\hfill
	\includegraphics[width=.5\textwidth]{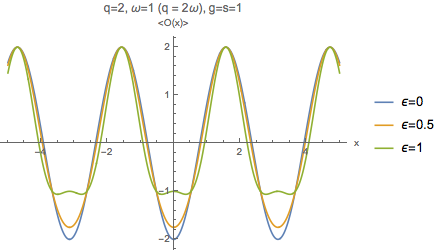}\hfill \\
	\includegraphics[width=.5\textwidth]{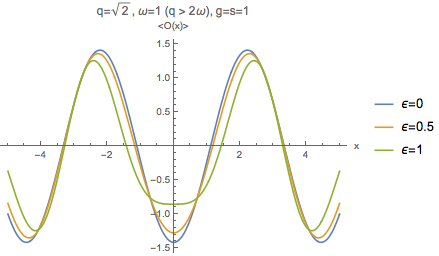}
	\caption{Expectation value of the boundary operator $\mathcal{O}$ Eq.~(\ref{deltaoscx}) for source $s(x)=\cos{qx}$ and inhomogeneous geometry $g(x)=\cos{\omega x}$ with different configurations of $(q,\omega)$}
	\label{fig:3}
	\end{figure}
		\item $g(x)=\sum\limits_{n=1}^{N}A_n \cos{\omega_n x}$. This polychromatic case is a direct extension of the above. It can be obtained by taking $N\to 2N$ and taking $A_n \to \frac{1}{2}A_n$ for $n=1,\dots, 2N$, $\omega_n \to \omega_n$ for $n=1,\dots, N$ and finally $\omega_n \to -\omega_n$ for $n=N+1,\dots, 2N$. By carefully splitting the sum and rearranging the terms, we get	$\Delta(x) = \Delta_{>}(x)+\Delta_{<}(x)$ with
		\begin{align}
		\Delta_{>}(x)=- \frac{\epsilon^2}{8}\sum\limits_{n=1}^{N}\sum\limits_{m=1}^{N}A_n A_m & \left[\frac{|\omega_n+\omega_m+q|-|q|}{\omega_n+\omega_m}\frac{\cos\left((\omega_n+\omega_m+q)x\right)}{\cos{qx}}\right.\notag\\
		&\left.-\frac{|\omega_n+\omega_m-q|-|q|}{\omega_n+\omega_m}\frac{\cos\left((\omega_n+\omega_m-q)x\right)}{\cos{qx}}\right],\notag\\
		\Delta_{<}(x)=-\frac{\epsilon^2}{8}\sum\limits_{n=1}^{N}\sum\limits_{m=1}^{N}A_n A_m & \left[\frac{|\omega_n-\omega_m+q|-|q|}{\omega_n-\omega_m}\frac{\cos\left((\omega_n-\omega_m+q)x\right)}{\cos{qx}}\right.\notag\\
		&\left.-\frac{|\omega_n-\omega_m-q|-|q|}{\omega_n-\omega_m}\frac{\cos\left((\omega_n-\omega_m-q)x\right)}{\cos{qx}}\right].
		\end{align}
		Note that since both terms are symmetric under the exchange $n\leftrightarrow m$, we can split the sum into a diagonal and an upper diagonal part, $\sum\limits_{n,m} = \sum\limits_{1\leq n\leq N}+2\sum\limits_{1\leq m<n \leq N}$. The expressions can suggest that the diagonal part in $\Delta_{<}$ diverges. However this is not the case, since the denominators also vanish, and we have to take the limit carefully.
		
		This polychromatic case is of particular interest since it can be used to simulate a discrete representation of disorder, as discussed in Appendix \ref{app:A}. Defining $\Delta\omega = \frac{\pi}{N a}$ for some constant lattice spacing $a>0$ and letting $A_n = \sqrt{\Delta \omega}$, $\omega_n=n\Delta\omega$ for $n=1,\dots,N$ and adding random phases i.i.d. uniformly $\gamma \in [0,2\pi)$ in the cosine arguments, $g(x)$ become a discrete representation of the Gaussian random process for large $N\gg 1$. As we will discuss in the next section, continuous disorder is more subtle, and we have to take the average at an early stage to make progress. Moreover, it is harder to implement it numerically since it can usually have discontinuous derivatives. For these reasons, this implementation is a useful representation and is specially suited for holography calculations, having been used in many previous works in the literature \cite{Arean2014,Hartnoll2014a, Garcia-Garcia2016}. 
		In this case all modes $\omega_n$ are positive, and we can assume without loss of generality that $q>0$. For convenience, we separate the sum in the correction in a diagonal and a non-diagonal part, $\Delta = \Delta_{\text{d}}+\Delta_{\text{nd}}$ with
	\begin{align}
	\label{deltadis}
	\Delta_{\text{d}}(x)=-\frac{\epsilon^2}{4}\sum\limits_{n=1}^{N}A_n^2\left[2-\right.&\left.\frac{|2\omega_n-q|-|q|}{2\omega_n}\frac{\cos\left((2\omega_n-q)x+2\gamma_n\right)}{\cos{qx}}+\right. \notag\\
	&+\left.\frac{\cos\left((2\omega_n+q)x+2\gamma_n\right)}{\cos{qx}}\right]\notag\\
	\end{align}
	
	\begin{align}
	\Delta_{\text{nd}}=-\frac{\epsilon^2}{4}\sum\limits_{1\leq m< n\leq N}\frac{A_n A_m}{\cos{qx}}\Big[&\cos\left((\omega^{-}_{nm}+q)x+\gamma_n-\gamma_m\right)+\notag\\
	&+\cos\left((\omega^{+}_{nm}+q)x+\gamma_n+\gamma_m\right)+\notag\\
	&-\frac{|\omega^{+}_{nm}-q|-|q|}{\omega^{+}_{nm}}\cos\left((\omega^{+}_{nm}-q)x+\gamma_n+\gamma_m\right)+\notag\\
	&-\frac{|\omega^{-}_{nm}-q|-|q|}{\omega^{-}_{nm}}\cos\left((\omega^{-}_{nm}-q)x+\gamma_n-\gamma_m\right)\Big].
	\end{align}
	\noindent where for convenience we abbreviated $\omega^{\pm}_{nm} = \omega_n \pm \omega_m$. Note that only the constant term in the diagonal part survives averaging over the i.i.d. phases $\gamma_n$ as discussed following Eq.\eqref{eq:appA:average} in Appendix \ref{app:A}. As expected, we reproduce the results for the constant geometry in Section \ref{sec:constant} on average. In Fig.\ref{fig:4} we plot the expectation value of the dual operator for $N=10$ modes and fixed source $q=\sqrt{2}$ for a fixed realization of phases. 
	\begin{figure}[htp]
	\centering
	\includegraphics[width=.5\textwidth]{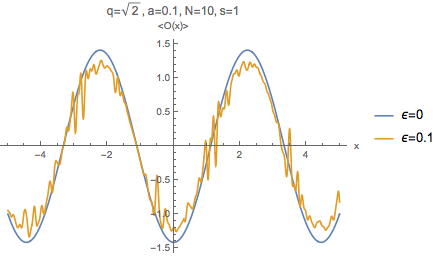}\hfill
	\includegraphics[width=.5\textwidth]{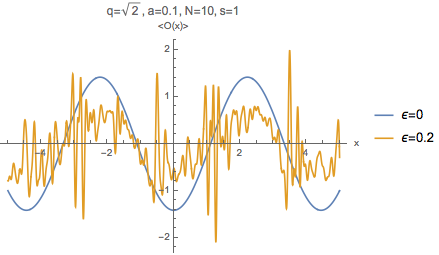}
	\caption{Expectation value of the boundary operator $\mathcal{O}$ Eq.~(\ref{deltadis}) for source $s(x)=\cos{\sqrt{2}x}$ and discrete implementation of a disordered geometry $g(x)=\sum\limits_{n=1}^{N}A_n \cos{\left(\omega_n x+\gamma_n\right)}$ for different amplitudes of disorder.}
	\label{fig:4}
	\end{figure}
		
	\end{itemize}
	
	\item[Delta source and two-point function] We now consider the case $s_k = 1$ that leads to the corrections to the boundary-to-bulk propagator $K$. The integral Eq.\eqref{eq:2ndosci} can be done explicitly,
	\begin{align}
K_{(2)}(\rho;x)	&= \frac{\rho^{1/4}}{2}\sum\limits_{n,m=1}^{N}\frac{A_{n}A_{m}}{\omega_n+\omega_m}\int_{\mathbb{R}}\frac{\dd k}{2\pi} (k-\omega_n-\omega_m) \left(e^{-|k-\omega_n-\omega_m|\sqrt{\rho}}-e^{-|k|\sqrt{\rho}}\right)e^{ikx},\notag\\
&=\frac{1}{2\pi}\frac{\rho^{3/4}}{x^2+\rho}\sum\limits_{n,m=1}^{N}A_{n}A_{m}\left[1-\frac{4 e^{i\frac{(\omega_n+\omega_m)x}{2}}\sin{\frac{(\omega_n+\omega_m)x}{2}}}{\omega_n+\omega_m}\frac{x}{x^2+\rho}\right].
\end{align}
Recall that the two-point function of the dual boundary operator is obtained by evaluating the boundary-to-bulk propagator at the boundary. Recalling that the zeroth order result is given by $\frac{1}{\pi x^2}$, this yields to a relative correction to two-point function given by
\begin{align}
\frac{\delta\langle\mathcal{O}(x)\mathcal{O}(0)\rangle}{\langle\mathcal{O}(x)\mathcal{O}(0)\rangle_{(0)}} = \frac{\epsilon^2}{2}\sum\limits_{n,m=1}^{N}A_{n}A_{m}\left[1-\frac{4 e^{i\frac{(\omega_n+\omega_m)x}{2}}\sin{\frac{(\omega_n+\omega_m)x}{2}}}{(\omega_n+\omega_m)x}\right].
\end{align}
Note that both terms are symmetric over $n \leftrightarrow m$. Thus we can split the sum into a diagonal term and a term where $n < m$,
\begin{align}
\frac{\delta\langle\mathcal{O}(x)\mathcal{O}(0)\rangle}{\langle\mathcal{O}(x)\mathcal{O}(0)\rangle_{(0)}} =&\frac{\epsilon^2}{2}\sum\limits_{n=1}^{N}A_{n}^2\left(1-\frac{2 e^{i\omega_n x}\sin{\omega_n x}}{\omega_n x}\right)+\notag\\
&+\epsilon^2\sum\limits_{1\leq n<m\leq N}A_{n}A_{m}\left(1-\frac{4 e^{i\frac{(\omega_n+\omega_m)x}{2}}\sin{\frac{(\omega_n+\omega_m)x}{2}}}{(\omega_n+\omega_m)x}\right)\bigg].
\end{align}
Consider the following interesting examples,
\begin{itemize}
\item $g(x) = g \cos{\omega x}$.

In this case we simply have
\begin{align}
\label{twoposci}
\frac{\delta\langle\mathcal{O}(x)\mathcal{O}(0)\rangle}{\langle\mathcal{O}(x)\mathcal{O}(0)\rangle_{(0)}} = -\epsilon^2\frac{g^2}{2} \frac{\sin{2\omega x}}{2\omega x},
\end{align}
\noindent which is proportional to a sinc function. Recall that we are looking at correlations of the point $x$ with the origin. Thus the correction induced by the oscillating geometry suppress correlations of points closer to the origin. Note that for $\omega \gg 1$, the peak is sharper, while for $\omega \ll 1$ it is broader. Thus the bigger the wavelength of the inhomogeneous geometry, the less is the correction localized at $x=0$, see Fig.\ref{fig:5}. We can check that the limit $\omega \to 0$ reduces to the constant case discussed in Section \ref{sec:constant}. 

	\begin{figure}[htp]
	\centering
	\includegraphics[width=.5\textwidth]{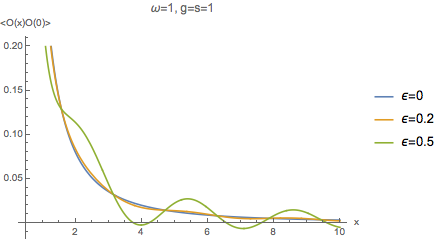}\hfill
	\includegraphics[width=.5\textwidth]{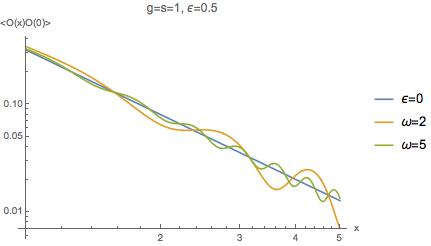}
	\caption{Two-point function $\langle\mathcal{O}(x)\mathcal{O}(0)\rangle = \langle\mathcal{O}(x)\mathcal{O}(0)\rangle_{(0)} + \epsilon^2 \delta\langle\mathcal{O}(x)\mathcal{O}(0)\rangle$ for the oscillating geometry $g_{tx} = \epsilon \cos{\omega x}$, from Eq.\eqref{twoposci}. On the left, we fix the frequency and plot different amplitudes of $\epsilon$, on the right we fix $\epsilon$ and plot different frequencies, in a log-log scale.}
	\label{fig:5}
	\end{figure}

\item $g(x) = \sum\limits_{n=1}^{N}A_n \cos{\left(\omega_n x + \gamma_n\right)}$.

The result above can be easily generalized for many modes. Following the discussion in for the single cosine source, we have:
\begin{align}
\frac{\delta\langle\mathcal{O}(x)\mathcal{O}(0)\rangle}{\langle\mathcal{O}(x)\mathcal{O}(0)\rangle_{(0)}}  &= \frac{\epsilon^2}{4}\sum\limits_{n=1}^{N}\sum\limits_{m=1}^{N}A_n A_m \left[1-\frac{\sin\left(\omega^{+}_{nm}x+\gamma_n-\gamma_m\right)}{\omega_{nm}^+ x}-\frac{\sin\left(\omega_{nm}^- x+\gamma_n-\gamma_m\right)}{\omega_{nm}^-x} \right],\notag \\
&=	\frac{\epsilon^2}{4}\sum\limits_{n=1}^{N}A_n^2\left[1-\frac{\sin(2\omega_n x+2\gamma_n)}{2\omega_n x}\right]\notag\\
 &+\frac{\epsilon^2}{2}\sum\limits_{1\leq m<n \leq N}A_n A_m \left[1-\frac{\sin\left(\omega^{+}_{nm}x+\gamma_n-\gamma_m\right)}{\omega_{nm}^+ x}-\frac{\sin\left(\omega_{nm}^- x+\gamma_n-\gamma_m\right)}{\omega_{nm}^-x} \right].
\end{align}
Interestingly this discrete random process has mean zero.

\end{itemize}
\end{description}

\subsubsection{Disordered Geometries}
\label{sec:disorderedgeom}
We now study the case where $g(x)$ is a centered random Gaussian field. As discussed in Appendix \ref{app:A}, we can decompose $g$ in its spectral modes $g(x) = \int_{\mathbb{R}}\frac{\dd k}{2\pi} g_k e^{ikx}$, with $g_k$ also given by a centered Gaussian process with $\mathbb{E}[g_k]=0$ and $\mathbb{E}[g_k g_l] = g^2\delta(k-l)$. Equation \eqref{eq:2ndFT} becomes a stochastic integro-differential equation, and is still intractable. However since it depends on the square of the geometry we can consider its non-trivial average. Defining $\bar{f}_k = \mathbb{E}[f_k]$ and averaging over $g$ yield
\begin{align}
\label{eq:2nddisorder}
4\rho^2 \bar{f}_{k}''-\left(\rho k^2-\frac{3}{4}\right)\bar{f}_k &=-\rho^{5/4}g^2\int_{\mathbb{R}}\dd q~(k-2q)(k-q) e^{-|k-2q|\sqrt{\rho}} s_{k-2q},\notag\\
&=-\frac{\rho^{5/4}g^2}{4}\int_{\mathbb{R}}\dd l~l(k+l) e^{-|l|\sqrt{\rho}} s_{l}.
\end{align}
As before, the constant source case give a trivial result. We now analyze non-trivial settings.
\begin{description}
	\item[Plane wave source] Let $s(x)=s e^{iq x}$ with $q>0$. Integrating Eq.\eqref{eq:2nddisorder} gives
	\begin{align}
		4\rho^2 \bar{f}_{k}''-\left(\rho k^2-\frac{3}{4}\right)\bar{f}_k &=-\frac{1}{4}\rho^{5/4}g^2 s q(k+q) e^{-|q|\sqrt{\rho}}.
	\end{align}
	The solution satisfying boundary conditions Eq.\eqref{2ndbc1} and \eqref{2ndbc2} is given by
	\begin{align}
	\bar{f}_{k}(\rho) = \frac{1}{4}\rho^{1/4}g^2 s\frac{q}{k-q}\left[e^{-|q|\sqrt{\rho}}-e^{-|k|\sqrt{\rho}} \right]	.
	\end{align}
	Note that the case $k=q$ can be treated by taking the limit of the above. Fourier transforming back,
	\begin{align}
	\bar{\psi}_{(2)}(\rho, x) = \int_{\mathbb{R}}\frac{\dd k}{2\pi} e^{ikx} \bar{f}_{k}(\rho) = \frac{1}{4}\rho^{1/4} g^2 s e^{iqx}\left[e^{-|q|\sqrt{\rho}}\text{E}_{1}\left(iqx-q\sqrt{\rho}\right)-e^{|q|\sqrt{\rho}}\text{E}_{1}\left(iqx+q\sqrt{\rho}\right)\right].
	\end{align}
	The correction to the one-point function is given by
	\begin{align}
	\overline{\Delta(x)} = \frac{\overline{\delta\langle\mathcal{O}\rangle}}{\overline{\langle\mathcal{O}\rangle}_0} =\frac{\epsilon^2}{2}g^2 \left[\frac{i e^{-iqx}}{qx} + \text{E}_{1}(iqx)\right].
	\end{align}
	
	\item[Oscillating source] Let $s(x)=s\cos{qx}$. As before, by linearity of Eq.\eqref{eq:2nddisorder}, we can build the oscillating case by summing two plane waves. Using that $\overline{\text{E}_1(z)} = \text{E}_1(\bar{z})$ we can simplify $\text{E}_1(iqx)+\text{E}_1(-iqx) = 2\text{Re}\left[\text{E}_1(iqx)\right] = -2\text{Ci}(iqx)$, where $\text{Ci}(x)$ is the cosine exponential function\footnote{Note that here the bars refer to the complex conjugate, and not to the average.}. This leads to
	\begin{align}
	\overline{\Delta(x)} = -\epsilon^2 g^2 \left[\frac{\sin{qx}}{qx}+\text{Ci}(qx)\right]	.
	\end{align}

	\item[Delta source and two-point function] Consider $s_k=1$. Integrating the right-hand side of Eq.\eqref{eq:2nddisorder},
		\begin{align}
		\label{randeq}
		4\rho^2 \bar{f}_{k}''-\rho k^2\bar{f}_k+\frac{3}{4}\bar{f}_k=-\rho^{-1/4}g^2.
		\end{align}
		The regular solution at $\rho=\infty$ for the above is given by
		
\begin{align}
\label{solrand}
f_{k}(\rho)=a_k\rho^{1/4}e^{-|k|\sqrt{\rho}}-\frac{g^2}{2\rho^{1/4}}+\frac{g^2}{4}\rho^{1/4}ke^{-k\sqrt{\rho}}\text{Ei}(k\sqrt{\rho})-\frac{g^2}{4}\rho^{1/4}k e^{k\sqrt{\rho}}\text{Ei}(-k\sqrt{\rho}),
\end{align}
\noindent where $\text{Ei}(x)=\int_{-x}^{\infty}t^{-1}e^{-t}\dd t$ is the exponential integral functions. Close to the boundary $\rho=0$ we have
\begin{align}
f_{k}\underset{\rho=0}{\sim}-\frac{g^2}{2\rho^{1/4}}+a_k\rho^{1/4}+\frac{g^2}{2}\left(\gamma-1-\log{k\sqrt{\rho}}\right)	\rho^{3/4}+\dots ~.
\end{align}
Therefore in order to preserve the zeroth order boundary condition we need to set $a_k=0$. Note the appearance of two divergences: one proportional to $\rho^{-1/4}$ and the other proportional to $\log{k\sqrt{\rho}}$. As we will see next, they recombine when taking the Fourier transform and lead to a finite result. 

Using the following Fourier transform that can be calculated from the definition of $\text{Ei}(x)$,
\begin{align}
\int_{-\infty}^{\infty}\frac{\dd k}{2\pi} e^{-a k} \text{Ei}(ak)e^{ikx} = -\frac{1}{2}\frac{|x|-i a~\text{sgn}(x)}{x^2+a^2}, && a>0 	
\end{align}
\noindent we can deduce that
\begin{align}
\int_{-\infty}^{\infty}\frac{\dd k}{2\pi} \left[e^{-a k} \text{Ei}(ak)-e^{a k}\text{Ei}(-ak)\right]e^{ikx}	 = \frac{ia~\text{sgn}(x)}{x^2+a^2}.
\end{align}
Now using that $\mathcal{F}[ik f_k] = \partial_x f(x)$, 
\begin{align}
\int_{-\infty}^{\infty}\frac{\dd k}{2\pi} \frac{k}{4}\left[e^{-a k} \text{Ei}(ak)-e^{a k}\text{Ei}(-ak)\right]e^{ikx}= -\frac{1}{2}\frac{a^2|x|}{(a^2+x^2)^2}+\frac{1}{2}\delta(x).
\end{align}
Letting $a=\sqrt{\rho}>0$ and taking into account the constant term leads to
\begin{align}
K_{(2)}(\rho;x) = \int_{\mathbb{R}}\frac{\dd k}{2\pi} f_{k}(\rho) e^{ikx} = -\frac{g^2}{2}\frac{\rho^{3/4}|x|}{(\rho+x^2)^2}.
\end{align}
Note that the first term in Eq.\eqref{solrand} leads to a delta function with opposite sign that exactly cancels the one coming from the derivative of the sign function. As discussed in Appendix \ref{app:C}, this result is finite close to the boundary
\begin{align}
K_{(2)}(\rho;x) \underset{\rho=0}{\sim} -\frac{g^2}{2}\frac{1}{|x|^3}+\dots ~.
\end{align}
Therefore the correction to the two-point function can be written as
\begin{align}\label{distwo}
\langle\mathcal{O}(x)\mathcal{O}(0)\rangle =\left(1-\frac{\epsilon^2}{2}\frac{\pi}{|x|}\right) \frac{1}{\pi}\frac{1}{x^2},
\end{align}
\noindent 
%or in the notation from the previous sections, $\Delta(x) = -\frac{\epsilon^2}{2}\pi |x|^{-1}$. 
The effect of disorder is similar to the inhomogeneous case discussed in Section \ref{sec:oscillating}. For points which are close together ($x=0$) we have $|\frac{\delta\langle\mathcal{O}(x)\mathcal{O}(0)\rangle}{\langle\mathcal{O}(x)\mathcal{O}(0)\rangle_{(0)}}|\gg 1$ which indicates the breaking of the perturbative analysis. The minus sign of the correction indicates a suppression of the zeroth order result, see Fig.~\ref{fig:2ptdisorder}. 
We shall see that the exponent for which the correction blows up as $x\to 0$ depends on both the mass of the scalar field 
(here fixed to $m^2 = -3/4$) and on the type of disorder. In the following subsection, we will relax these 
conditions and explore the dependence of our results on these parameters. 

	\begin{figure}[htp]
	\centering
	\includegraphics[width=.5\textwidth]{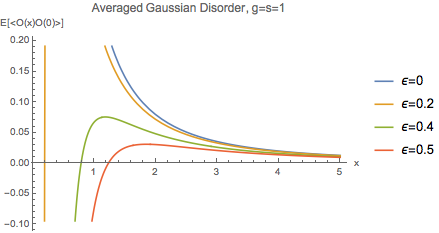}\hfill
	\includegraphics[width=.5\textwidth]{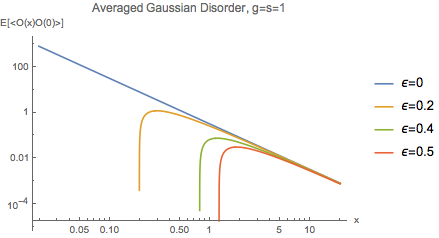}
	\caption{Averaged two-point function Eq.~(\ref{distwo}) for oscillating geometry disordered Gaussian geometry for different disorder strength $\epsilon$. The plot on the right is the same as the one in the left, but in log-log scale.}
	\label{fig:2ptdisorder}
	\end{figure}
\end{description}

\subsubsection{Comments on other masses and correlated disorder}
In this subsection we discuss how the results for the averaged two-point function generalize to different masses and correlated disorder. 

We start by considering different masses. It is easy to check that for a generic mass parametrized by $\nu^2 = m^2+1$ the zeroth order solution of Eq.\eqref{eq:0eq} that satisfies the boundary conditions Eqs.\eqref{bc1},\eqref{bc2} is given by,
\begin{align}
\label{eq:zeroth:genmass}
f_{k}(\rho) = \frac{|k|^{\nu}}{2^{\nu-1}\Gamma(\nu)}K_{\nu}\left(|k|\sqrt{\rho}\right) s_k.
\end{align}
The boundary-to-bulk propagator is, as before, obtained by setting $s_k=1$ and computing the Fourier transform, giving 
\begin{align}
K_{(0)}(\rho,x) = \frac{\rho^{1/2}}{2^{\nu-1}\Gamma(\nu)}\int_{\mathbb{R}}\frac{\dd k}{2\pi} e^{ikx} |k|^{\nu}K_\nu\left(|k|\sqrt{\rho}\right) = \frac{\Gamma\left(\nu+\frac{1}{2}\right)}{\sqrt{\pi}\Gamma(\nu)}	\frac{\rho^{\frac{\nu+1}{2}}}{\left(x^2+\rho\right)^{\nu+\frac{1}{2}}}.
\end{align}
Taking the near boundary limit, we obtain the expected dual two-point function,
\begin{align}
\langle\mathcal{O}(x)\mathcal{O}(0)\rangle_{(0)}=\lim\limits_{\rho\to 0}\rho^{-\frac{1+\nu}{2}}K_{(0)}(\rho,x)	= \frac{\Gamma\left(\nu+\frac{1}{2}\right)}{\sqrt{\pi}\Gamma(\nu)}	\frac{1}{|x|^{2\nu+1}} = \frac{\Gamma\left(\nu+\frac{1}{2}\right)}{\sqrt{\pi}\Gamma(\nu)}	\frac{1}{|x|^{2\Delta_{+}-1}},
\end{align}
\noindent which the the expected result for the two-point function of a conformal operator with mass dimension $\Delta_{+}$. The minus one factor is, as before, due to the time dimensional reduction. The calculation of corrections induced by the disordered geometry are exactly as before, with the only difference that we use the general solution above as a source in Eq.\eqref{eq:2ndorder}. Averaging the right-hand side, we get the general equation
\begin{align}
4\rho^2 \bar{f}_{k}''-\left(\rho k^2+\nu^2-1\right) \bar{f}_k  = -\frac{\sqrt{\pi}\Gamma\left(\nu+\frac{3}{2}\right)}{\Gamma(\nu)}\rho^{-
\nu/2},
\end{align}
\noindent which reduces to Eq.\eqref{randeq} for $\nu=1/2$. The general solution for the above is a combination of hypergeometric functions. So next we consider other values of $\nu$ for which the calculations are less cumbersome. Take for instance $\nu = 3/2$. In this case the solution satisfying the boundary conditions is given by
\begin{align}
\label{eq:2nd:fouriersol}
f_{k}(\rho) = -\frac{1}{\rho^{3/4}}+\frac{k}{4\rho^{1/4}}\left[e^{-k\sqrt{\rho}}\left(1+k\sqrt{\rho}\right)\text{Ei}\left(k\sqrt{\rho}\right)-e^{k\sqrt{\rho}}\left(1-k\sqrt{\rho}\right)\text{Ei}\left(-k\sqrt{\rho}\right)\right].
\end{align}
The Fourier transform of the above can be computed exactly in the same way as in Section \ref{sec:disorderedgeom} and is given by
\begin{align}
K_{(2)}=-\rho^{5/4}\frac{2|x|}{(\rho+x^2)^3}	, && \nu = 3/2.
\end{align}
Noting that for $\nu = 3/2$ we have $\Delta_{+
} = 1+\nu =5/2$, the dual two-point function is given by
\begin{align}
\langle \mathcal{O}(x)\mathcal{O}(0)\rangle = \frac{2}{\pi}\left(1-\frac{\epsilon^2}{2}\frac{\pi}{|x|}\right)\frac{1}{|x|^4}, && \nu = 3/2.
\end{align}
This result follow exactly the one for $\nu=1/2$, with the only difference that now the zeroth order result has a different mass dimension. It is not hard to check that the same is true for $\nu=5/2$, where the averaged two-point function is given by
\begin{align}
\langle \mathcal{O}(x)\mathcal{O}(0)\rangle = \frac{8}{3\pi}\left(1-\frac{\epsilon^2}{2}\frac{3\pi}{|x|}\right)\frac{1}{|x|^6}, && \nu = 5/2,
\end{align}
\noindent and now $2\Delta_{+} -1 = 6$. Although we have not manage to prove the general result, it seems that white noise always induce the same relative corrections in the two-point function. As we will discuss next, this is not completely surprising. For higher masses the dual operator is a more relevant deformation, but we are not changing the mass dimension of disorder. 

To see this explicitly, note that the metric element $g_{tx}$ should be dimensionless. Since we are imposing $g_{tx} = \epsilon \int\frac{\dd k}{2\pi}e^{ikx}f_{k}$, we must have $[\epsilon] = -1-[f_{k}]$. On the other hand, for Gaussian white noise $\mathbb{E}[f_kf_l] = \delta(k-l)$ and thus $2[f_k] = -1$. Therefore in this case we must have $[\epsilon] = -1+1/2 = -1/2$, i.e. Gaussian white noise is irrelevant. To change the effective mass dimension of disorder, we can consider correlated disorder $\mathbb{E}[f_kf_l] = \sigma^2_k\delta(k-l)$ with $\sigma_{k}^2=|k|^{\alpha}$, $\alpha\in \mathbb{R}$. Note that since $\sigma_k^2$ is a variance, it needs to be a positive definite function. Generalising the previous discussion, this gives
\begin{align}
[\epsilon] = -\frac{\alpha+1}{2}.	
\end{align}
This is precisely what we found in the previous section: corrections to the two-point function decay faster in the IR limit $|x|\to\infty$ than the leading order result. Choosing $\alpha>0$ will only make disorder more irrelevant. It is easy to check that for $\alpha>0$, we get subleading powers of $\rho$ which do not contribute to the two-point function.

We now explore the case $\alpha = -2$, when disorder becomes relevant. For simplicity, lets consider again $\nu=1/2$. In this case, the integral in the right-hand side of Eq.\eqref{eq:2nddisorder} can be written as
\begin{align}
\int\dd q~ \frac{(k-2q)(k-q)}{q^2} e^{-|k-2q|\sqrt{\rho}} &=(2k\sqrt{\rho}+1)e^{k\sqrt{\rho}}E_{2}(k\rho)-(2k\sqrt{\rho}-1)e^{k\sqrt{\rho}}E_{2}(k\rho)-4,&\notag\\
&= 2\rho^{-1/2}+k\Big[(2k\sqrt{\rho}+3)e^{k\sqrt{\rho}}\text{Ei}(-k\sqrt{\rho})+\notag\\
&\hspace{6em}+(2k\sqrt{\rho}-3)e^{-k\sqrt{\rho}}\text{Ei}(-k\sqrt{\rho})\Big],
\end{align}
\noindent where $E_\alpha(z) = z^{\alpha-1} \int_{z}^{\infty}\dd t~ t^{-\alpha}e^{-t}$ is the generalised exponential integral function, which in the last line we related to the exponential integral through the following recursive relation
\begin{align}
p~E_{p+1}(z)+z~E_{p}(z) = e^{-z}	.
\end{align}	
\noindent together with $E_{1}(z)=-\text{Ei}(-z)$. The regular solution at the Poincar\'e horizon which does not modify the source will be given by
\begin{align}
f_{k}(\rho) &= -\rho^{3/4}-\frac{1}{2}\rho^{5/4}\left[(k\sqrt{\rho}+2)e^{k\sqrt{\rho}}\text{Ei}(-k\sqrt{\rho})-(k\sqrt{\rho}-2)e^{-k\sqrt{\rho}}\text{Ei}(k\sqrt{\rho})\right],	\notag\\
&=-\rho^{3/4}-\rho^{3/4} F_{+}(k\sqrt{\rho})+\frac{\rho^{5/4}k}{2} F_{-}(k\sqrt{\rho}),\\
&= -\rho^{3/4}\left[1+F_{+}(\omega)-\frac{\omega}{2}F_{-}(\omega)\right],
\end{align}
\noindent where in the last equality we defined $\omega = k\sqrt{\rho}$. We can compute the Fourier transform of the above in following the same recipes as discussed above. The boundary-to-bulk propagator is given by
\begin{align}
K_{(2)}(\rho, x)= -\rho^{3/4}\frac{x^2|x|}{(\rho+x^2)^2},
\end{align}
\noindent and the correction to the boundary two-point function thus given by
\begin{align}
\langle\mathcal{O}(x)\mathcal{O}(0) \rangle	= \frac{1}{\pi}\frac{1}{|x|^2}-\frac{\epsilon^2}{|x|} =\frac{1}{\pi}\left(1-\epsilon^2 \pi |x|\right)\frac{1}{|x|^2} .
\end{align}
Note that different from the results in the previous section, the correction induced by the correlated Gaussian disorder dominates the decay of the propagator at the IR, $|x|\to\infty$. For any fixed amplitude $\epsilon>0$ and for $|x| > \frac{1}{\epsilon^2\pi}$ the correction becomes more important than the zero order result, which signals a breakdown of perturbation theory.

%%%%%%%%%%%%%%%%%%%%%%%%%%%%%%%%%%%%%%%%
\section{Numerical Analysis}
\label{sec:nume}
%%%%%%%%%%%%%%%%%%%%%%%%%%%%%%%%%%%%%%%%

We have also performed a numerical analysis which allows us to confirm and extend our previous results beyond perturbation theory. 
Our first step is to compactify the interval of the radial coordinate. Following \cite{Hartnoll2014a}, we define $y$ via
\begin{equation}\label{rho y}
	\rho = \frac{y^2}{(1-y)^2} 
\end{equation}
\noindent In this new coordinate, the boundary is located at $y= 0$ and the Poincar\'e horizon at $y= 1$. We redefine the scalar as
\begin{equation}
	\psi = \frac{y^{1/2}}{(1-y)^{1/2}} \chi
\end{equation}
\noindent We can check that the near boundary behavior can be expressed in terms of $\chi$ as
\begin{equation}\label{chi series}
	\chi \approx \chi^{(0)} + y \langle O \rangle + \ldots
\end{equation}
Here $\chi^{(0)}$ and $\langle O \rangle$ correspond to the source and vev of the dual operator.
Comparing with the pure AdS solution given in \eqref{eq:0thhom}, we see that, at least for $g_{tx} =0$ modes
with non-zero momentum decay or grow exponentially at the horizon. After turning on $g_{tx}$, the near horizon 
asymptotics become harder to analyze, but by continuity with the pure AdS case we demand that the field vanishes there. 

We solve the equation of motion by discretizing it on a homogeneous grid along both the radial coordinate $y$ and the boundary coordinate
$x$, and solving the resulting matrix equation in Mathematica. Since most of the variability occurs along $x$, we use $N_x = 450$
grid points along this direction, and $N_y = 50$ for the radial direction. After solving the wave equation, we extract the vev 
by taking a radial derivative of the solution at the boundary, as seen in \eqref{chi series}. 

We compute approximations to the one and two-point functions, concentrating on the case of disordered geometries. We do so by 
employing the spectral representation of disorder discussed in Appendix \ref{discrete}. More specifically, we take $g_{tx}$ 
to be given Gaussian by using \eqref{dspectral} with $\sigma(k_n) = 1$ with suitable modifications described in more detail below.

%
%%%%%%%%%%%%%%%%%%%%%%%%%%%%%%%%%%%%%%%%%%%%%%%%%%%%%%%
\subsection{One-point correlation function}
%%%%%%%%%%%%%%%%%%%%%%%%%%%%%%%%%%%%%%%%%%%%%%%%%%%%%%%
%
We consider one-point function of the scalar in the presence of the source
\begin{equation}
	\chi^{(0)} = \cos (k_0 x).
\end{equation}
As explained above, the vev corresponds to the derivative of the field at the boundary, as given by \eqref{chi series}. 
Since the periodicities of the source and geometry need to fit in the same computational domain, the expansion
of the disordered geometry will contain terms $ \cos(2 n k_0 x + \gamma)$ with integer $n$, see 
Eq. \eqref{dspectral}. 
As seen in Section \ref{sec:oscillating}, these are potentially problematic since the commensurability of the source with the geometry induces extra 
near horizon divergences. To avoid this undesirable behaviour, we only consider cosines of odd momentum in the sum Eq.\eqref{dspectral}. 
We shall see that with this modification we can still obtain some generic features of disorder which match well with the perturbation 
theory results for small disorder amplitude, and extend them to higher values. 

\begin{figure}[htp]
\centering
\includegraphics[width=.5\textwidth]{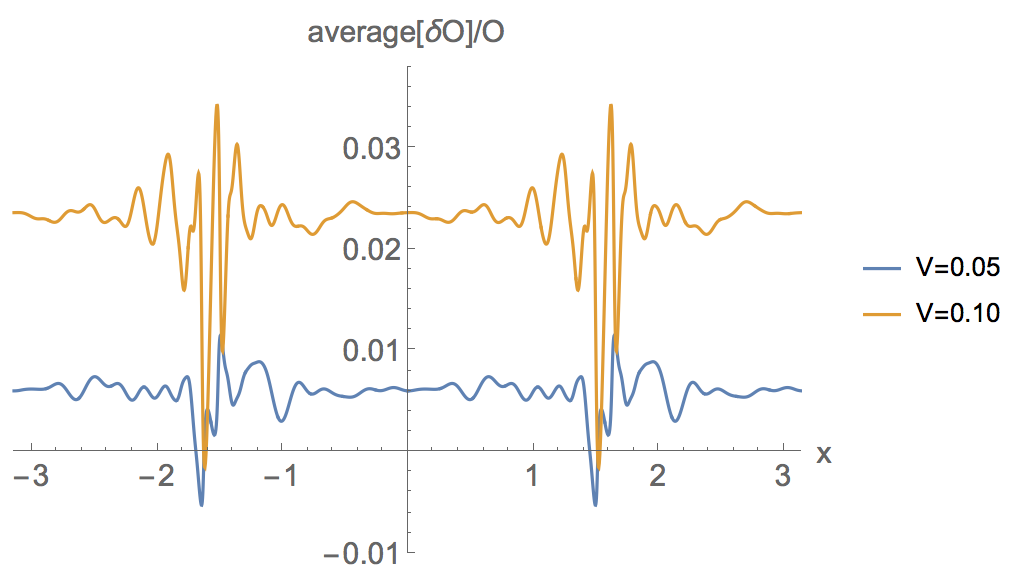}
\caption{Change of the one-point function due to the presence of disorder as a function of $x$ for different disorder amplitudes.
We plot $\overline{\delta \langle O \rangle}/ \langle O \rangle_0 = 1 - \eta$, where $\eta$ is defined in \eqref{def eta}.}
\label{fig:dOoverO}
\end{figure}

We extract the one-point function for $k_0 = 5$, and $N=20$ for varying values of the disorder amplitude $\bar V$ as defined 
in Appendix \ref{discrete}. For every $\bar V$, 
we generate a random geometry by providing random phases in Eq.\eqref{dspectral}. Once we have obtained a large number of them, 
we take the arithmetic average at each point $x$. 
We write the average vev as a suppression factor $\eta(\bar V, x)$ times the translational invariant result, given 
by $\langle O \rangle_0 := \langle O \rangle|_{{\bar V} = 0} = - k_0 \cos(k_0 x) $. Hence, we define $\eta$ by 
\begin{equation}\label{def eta}
	\overline{ \langle O \rangle }(x) = - \eta(\bar V, x) k_0 \cos(k_0 x)
\end{equation}
For small disorder amplitudes the $x$-dependence of $\eta(\bar V, x)$ is very mild. 
However, at larger amplitudes the $x$-dependence of $\eta$ becomes important. We show this in Fig. \ref{fig:dOoverO}
where we plot the average of $ \delta \langle O \rangle := \langle O \rangle_0 -  \langle O \rangle $, normalized by  $\langle O \rangle_0$,
as a function of $x$.

\begin{figure}[htp]
\centering
\includegraphics[width=.5\textwidth]{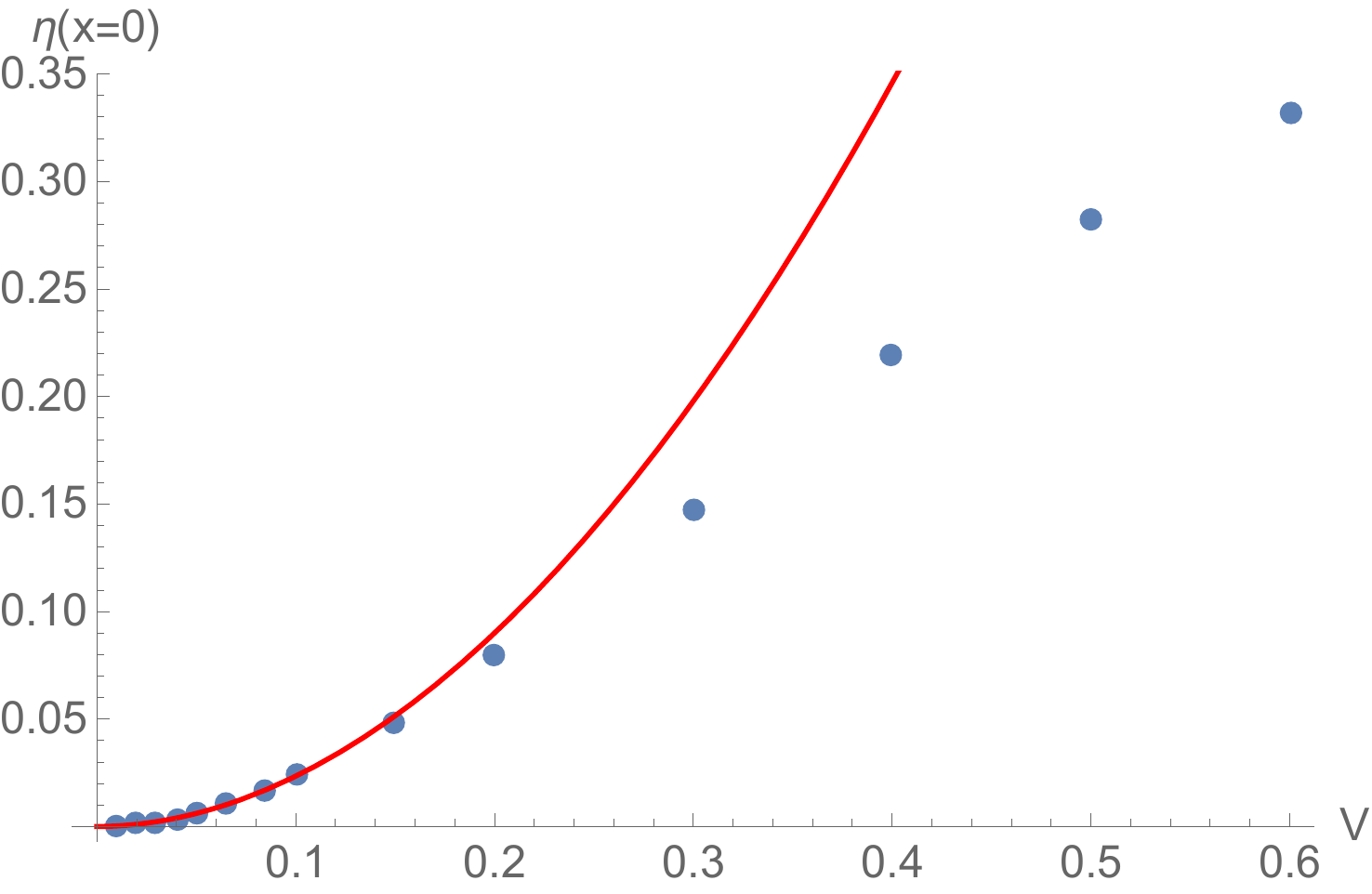}
\caption{Dependence of the suppression of the peak of the vev, $\eta$, defined in \eqref{def eta}, as a function of disorder. 
The solid red line is a fit we do for small $\bar V$, obtaining $\eta(x=0) \approx a \bar V^\gamma$ with $a = 2.05$, 
$\gamma = 1.94$. At higher $\bar V$ the exponent of the power law decreases.}
\label{fig:eta vs V}
\end{figure}

\begin{figure}[htp]
\centering
\includegraphics[width=.4\textwidth]{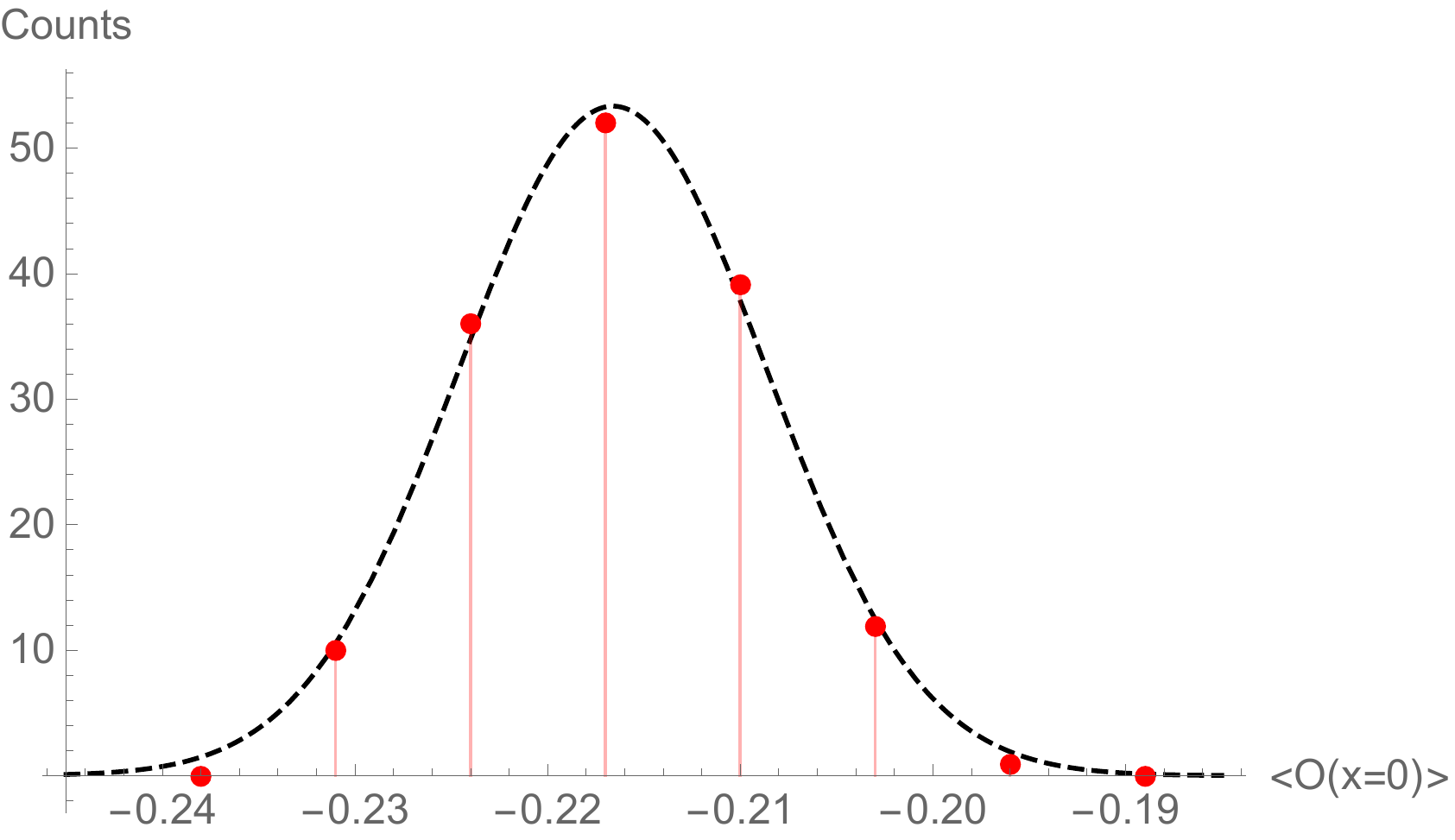}
\caption{Distribution of the value of the vev at the peak, $\langle O \rangle (x=0)  $ , for $\bar V=0.3$ for 150 runs. 
We fit this to a Gaussian, $f(x) = A \exp\left ( - (x- \mu)^2/(2 \sigma^2) \right)$,  with parameters $A = 53.3$, 
$\mu = -0.21$, $\sigma = 8 \times 10^{-3}$}
\label{fig:distrib peak 1pt}
\end{figure}

In order to estimate the overall suppression, we 
track the value at the peak $\eta(\bar{V}, x= 0)$. We show our results in Fig \ref{fig:eta vs V}.
For small $\bar V$, we observe that the averaged one-point function displays the quadratic behavior obtained in perturbation theory, although with different 
proportionality constant. At larger amplitudes, the power-law behavior becomes milder. 
Moreover, we are able to fit the distribution of values of the vev at the peak with a Gaussian centred at the average value, see 
Fig \ref{fig:distrib peak 1pt}. 

%%%%%%%%%%%%%%%%%%%%%%%%%%%%%%%%%%%%%%%%%%%%%%%%%%%%%%%
\subsection{Two-point correlation function}
%%%%%%%%%%%%%%%%%%%%%%%%%%%%%%%%%%%%%%%%%%%%%%%%%%%%%%%

We now obtain an approximation to the two-point function in the presence of a disordered geometry. 
In principle, this entails a highly expensive calculation which requires inserting arbitrary sources at different points
and taking the variation of the action with respect to them in the presence of a spatially dependent geometry. 
In order to gain some insight on the behavior, we consider the more tractable calculation corresponding to the 
two-point function $G(x,0)$, which as explained above can be obtained by inserting a delta function source at $y = 0$.
In order to regularize the delta function, we follow the strategy of \cite{Janik:2015oja}, i.e. we take as a source the boundary-to-bulk propagator
evaluated at a small cutoff. We stress that we need to take into account the fact that in our numerics the $x$ coordinate is periodic, 
which changes the form of the boundary-to-bulk propagator even in the absence of disorder $\bar V=0$.
In fact, it is easy to show that for a box of length $2 \pi /k_0 $ the boundary-to-bulk propagator is given by 
\begin{equation}
	K(x,0; y) =  \frac{ \sinh\left( - \frac{y}{1-y}\right) }{ \cos(k_0 x) - \cosh\left( - \frac{y}{1-y}\right) }
\end{equation}
Note that here $y$ refers to the radial variable introduced in Eq.\eqref{rho y}. 

\begin{figure}[htp]
\centering
\includegraphics[width=.4\textwidth]{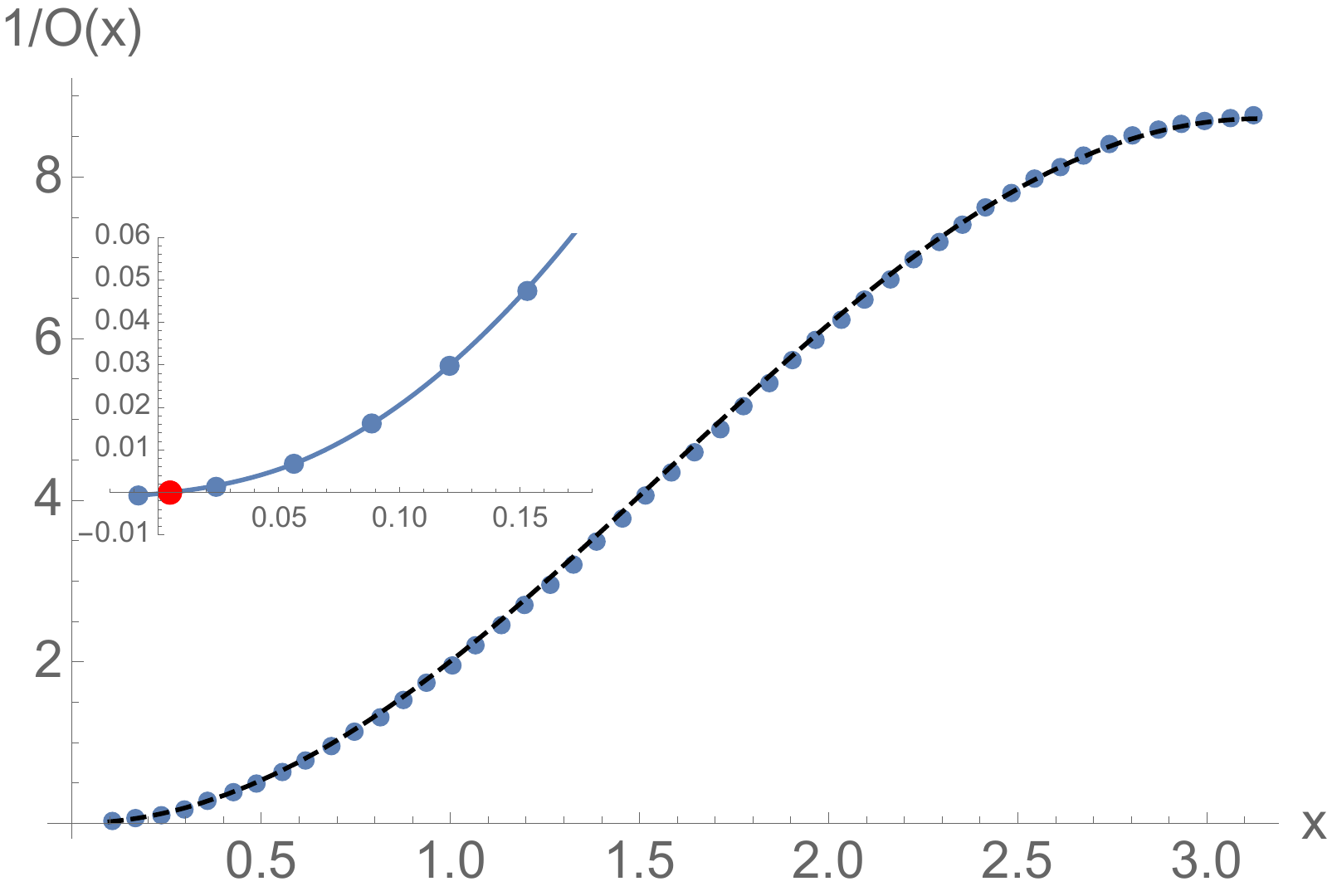}
\caption{We show our numerical result for the inverse of the two-point function $G(x,0)$ evaluated as the one-point function in the presence
of the source Eq.\eqref{delta reg 2pt} for $\delta = 0.01$ and $\bar V=0$. The blue dots represent the numerical data. 
The behavior away from the core deviates only by a multiplicative factor $\sim 1.09$ from the analytic result obtained 
without cutoff $\langle O(x)\rangle ^{-1} = k_0 (1 - \cos x)$. To illustrate this, we plot with the black dashed line the function 
$1.09 k_0 (1 - \cos x)$ showing good agreement with the numerics.
The inset shows the behaviour near $x = 0$. Here, the solid line shows interpolation of the 
numerical data, which indicates that two-point function acquires large, negative values near the core. The red dot marks the point where the
numerics diverge.}
\label{fig:2pt V=0}
\end{figure}

Therefore, we approximate the two-point function $G(x,0)$ by the one-point function obtained in the presence of the source 
\begin{equation}
\label{delta reg 2pt}
	\chi^{(0)} (x) = K(x,0; \delta)
\end{equation}
\noindent at small $\delta$. To test our approximation scheme, we first derive the results for pure AdS$_3$,  
$\bar V = 0$. As expected, this approximation fails for $x \approx 0$. In particular, the so-obtained vevs become very large and negative for 
small enough $x$. However, the results near the edge of the computational domain $x = \pi$ are well-behaved, and match well 
with the analytic result, see Fig \ref{fig:2pt V=0}. Therefore, we will be able to extract meaningful results away from the cores in this 
region\footnote{Note that since we are using periodic boundary conditions, we cannot go infinitely far away from the cores
of the delta functions.}. 

The quantity of interest will be the ensemble average value of the one-point function in the presence of the source 
Eq.\eqref{delta reg 2pt}, $\overline{ \langle O \rangle }(x) |_{\bar V} $. Normalizing this by the corresponding one-point function at $\bar V=0$, 
we define
\begin{equation}
	\sigma(\bar V, x) = \frac{\overline{ \langle O \rangle }(x) |_{\bar V}  }{ \langle O \rangle|_{\bar V=0} }
\end{equation}

In order to capture the behavior away from the core, we take the spatial average in the interval $(2 \pi /3, \pi)$, which we denote by 
\begin{equation}\label{tilde s}
	\tilde \sigma (\bar V) = \frac{3}{\pi} \int_{2 \pi/3}^\pi \dd x~ \sigma(\bar V, x)
\end{equation}
This gives an estimate of the suppression of the two-point function in the presence of disorder. 

\begin{figure}[htp]
\centering
\includegraphics[width=.5\textwidth]{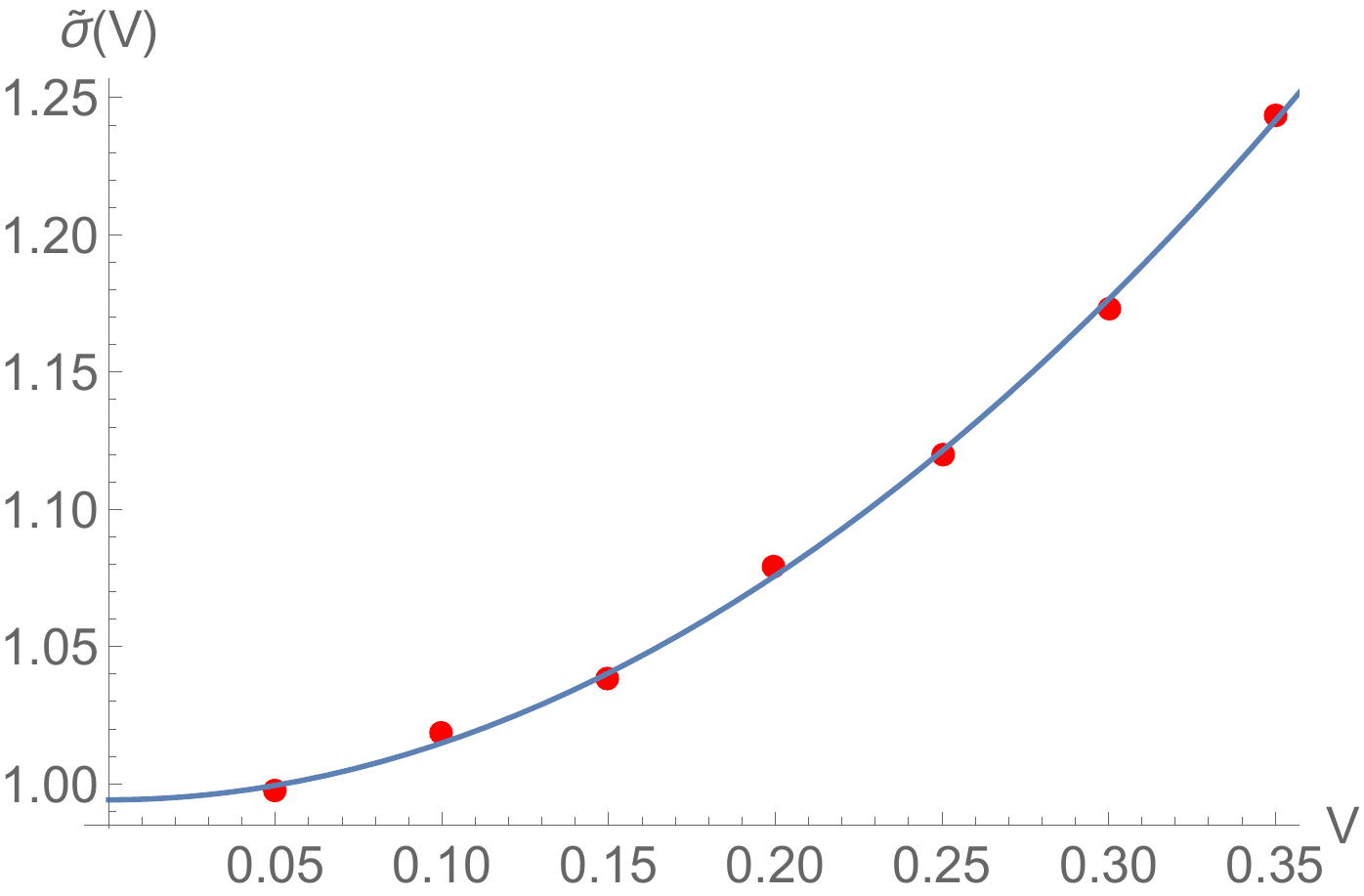}
\caption{Behavior of $\tilde \sigma$ defined in Eq.\eqref{tilde s} as a function of $\bar V$. The solid line corresponds to 
a fit with a model of the form $a + b \bar V^\gamma$ with $a = 0.99$, $b = 1.98$, $\gamma = 1.98$}
\label{fig:2pt suppresion}
\end{figure}

We plot our results for this quantity with $k_0 = 5$, $N=20$, $\delta = 0.01$ and varying $\bar V$ in Fig \ref{fig:2pt suppresion}. 
Once again, we obtain a quadratic dependence of the suppression with the disorder amplitude.  

%%%%%%%%%%%%%%%%%%%%%%%%%%%%%%%%%%%%%%%%
\section{Comparison with previous results in the literature}
\label{sec:comp}
%%%%%%%%%%%%%%%%%%%%%%%%%%%%%%%%%%%%%%%%
In the introduction, we reviewed different approaches to introduce disorder or spatial inhomogeneities in the holography literature. Now we discuss similarities and differences with the one introduced in this paper. 
A direct comparison is in general not possible in most cases. For instance the prediction for transport coefficients of \cite{Donos2013a,Donos2014b,Donos2014c,Donos2015,Donos2015a,Donos2017} requires the existence of a horizon and therefore are not applicable to our case. Even with no horizon, a comparison could be problematic because the assumption of a random chemical potential prevents, at least for a non-random background, any coherence effect. 
The addition of a random source, investigated in \cite{Fujita2008, Aharony2015}, is, to the best of our knowledge, not clearly connected to our approach. Instead of turning a source for the charge density, 
	in our model we introduce a source for the stress energy tensor. From the gravitational perspective, 
	we expect this to have a more significant effect because all fields couple to gravity. The equivalent field theory statement is that all 
	operators propagate on the fixed boundary geometry. Moreover the geometry in this approach is not random, so no coherence effects are expected to be observed.
It is an open question whether the observation of non-perturbative logarithmic corrections for marginal disorder in the two-point function, reported in \cite{Fujita2008, Aharony2015},
could occur in our setting. In order clarify these issues it would be necessary to 
carry out a full renormalization group analysis, beyond the scope of the paper, for the parameters for which the perturbative contribution from the disordered background becomes marginal. The result of such calculation would not only shed light on the existence of logarithmic corrections but also on the possible existence of a metal-insulator transition.

The approach closer to the one studied in the paper is maybe that of \cite{Adams2011,Hartnoll2014a,OKeeffe2015,Garcia-Garcia2016} where a spatially random chemical potential, or scalar, in the boundary, backreacts in the gravity background that becomes inhomogeneous as well. However there are still important differences. At least perturbatively, interesting coherence effects are strongly suppressed, even if no horizon is present, because the only independent source of randomness comes from the scalar or chemical potential whose profile in the boundary is fixed by boundary conditions. 

\section{Conclusions and outlook}
\label{sec:discu}
In this manuscript we have proposed a new approach to study disordered holographic field theories.
We have computed numerically and analytically corrections due to a weakly disordered gravity background in the one and two-point function of the scalar dual boundary operator for different choices of source and random component of the geometry $g_{tx}$ (constant, a superposition of plane waves and finally a Gaussian random function). The main results can be summarized as follows: 

\begin{itemize}
	\item A constant geometry induces a negative constant correction that decreases the overall amplitude of the scalar one and two-point functions.
	\item The corrections induced by an oscillating geometry have a richer behaviour, and depend on the interaction with the source. In the simple case where the source is also an oscillating function, the perturbative correction to the one-point function can be positive or negative, depending on the relative sign of the geometry and source frequencies. For instance, if the source frequency is much bigger than the geometry's, the correction is similar in form to the constant case outlined above. However if the source frequency is smaller than the geometry's, the correction flips sign, adding constructively to the zeroth order one-point function.

	\item An oscillating sinusoidal geometry induces an oscillating but decaying correction to the two-point function, with an envelope $\sim |x|^{-\alpha}$ that depends on the scalar mass.
	 
	\item The case in which the geometry is a superposition of oscillating modes is given, by linearity, by the sum of the single frequency results mentioned above. 
	
	\item We have identified coherence effects between a sinusoidal source and the weakly random geometry introduced by a spectral decomposition. In certain region of parameters, the one-point function, which is sinusoidal, in the absence of disorder, becomes completely random even in the limit in which perturbation theory applies.
	
	\item The averaged corrections to the two-point function in the presence of a delta source and a weakly random, Gaussian distributed, gravity background, is negative and, for large distances, decays as a power-law with an exponent that depends on the type of disorder and the scalar mass. We have identified a range or parameters for which perturbation theory breaks down, as the power-law decay is slower than in the non-random case. This suggest an instability to a novel disorder driven fixed point which could eventually lead to a metal-insulator transition in the system.  
\end{itemize}

Finally we mention some ideas for future research. First, we have only studied one instance of random geometry. Asymptotic AdS$_3$ solutions of Einstein's Equations have been classified in \cite{Skenderis2000}, and there are other families in which disorder could be introduced in a similar fashion. It would be interesting investigate whether introducing disorder in other metric components would lead to a similar universal behavior, and if not so, to identify the physics behind these differences. Second, since temperature tends to suppress coherence effects in disordered systems, in this work we have only studied holographic field theories at strictly zero temperature. But finite temperature solutions in three dimensions have also been classified \cite{Li2013}, and therefore our work could be also generalized in this direction. In particular, it would be interesting to study the low temperature regime and compare it with our results. Third, we have only considered Gaussian distributed disorder with delta-like correlations. However our formalism is easily generalizable to more general forms of disorder where correlations between points becomes important. Fourth, a renormalization group treatment, feasible for marginal random perturbations, would shed light on the existence, or not, of a novel non-trivial disordered driven fixed point which signals an instability toward a metal-insulator transition in the system. 

\section*{Acknowledgement}
We thank Ben Withers, Alexander Krikun and Aurelio Romero-Bermudez for illuminating discussions. The work of TA is supported by the ERC Advanced Grant GravBHs-692951. 
He also acknowledges the partial support of the Newton-Picarte Grant 20140053. BL is supported by a CAPES/COT grant No. 11469/13-17.
 
\bibliography{bibliography}
\bibliographystyle{JHEP}

\appendix
%%%%%%%%%%%%%%%%%%%%%%%%%%%%%%%%%%%%%%%%%%%%%%%%%%%%%%%%%%
\section{Notes on random fields}
%%%%%%%%%%%%%%%%%%%%%%%%%%%%%%%%%%%%%%%%%%%%%%%%%%%%%%%%%%
\label{app:A}

\subsection{Implementation}
Consider a random function $f:\mathbb{R}^d\to\mathbb{R}$. A useful trick to parametrise the randomness in $f$ is to work in the spectral representation (a.k.a. Fourier space)
\begin{align}
\label{spectral}
f(\vec{x}) = \int_{\mathbb{R}^{d}}\frac{{\dd[d] k}}{(2\pi)^d}	~f(\vec{k}) e^{i\vec{k}\cdot\vec{x}},
\end{align}
where $f(\vec{k})$ are random Fourier coefficients. In other words: we exchanged randomness in real space for randomness in Fourier space. Without loss of generality we can parametrise the Fourier coefficients $f(\vec{k})=a_{\vec{k}}+i b_{\vec{k}}$ where $a_{-\vec{k}}=a_{\vec{k}}$ and $b_{-\vec{k}}=-b_{\vec{k}}$ for reality of $f(\vec{x})$. We say $f$ is a \emph{Gaussian random field} when the Fourier coefficients $(a_{\vec{k}},b_{\vec{k}})$ are drawn from a Gaussian distribution
\begin{align}
	\label{distribution}
	P[f(\vec{k})]=P[a_{\vec{k}},b_{\vec{k}}] = \frac{1}{\pi\sigma_{\vec{k}}}e^{-\frac{a_{\vec{k}}^2+b_{\vec{k}}^2}{\sigma^2_{\vec{k}}}} = \frac{1}{\pi\sigma_{\vec{k}}}e^{-\frac{|f(\vec{k})|^2}{\sigma^2_{\vec{k}}}},
\end{align}
where $\sigma_{\vec{k}}$ is the standard deviation, and for simplicity we centred the distribution at zero. In other words, we have $\mathbb{E}[f(\vec{k})]=0$ and $\mathbb{E}[f(\vec{k})f(\vec{q})]=\sigma^2_{\vec{k}}\delta(\vec{k}+\vec{q})$. 
It is important for the distribution to be normalised:
\begin{align}
	\label{normalisation}
	\int\text{D}f(\vec{k}) ~P[f(\vec{k})]= \int_{-\infty}^{\infty}\dd a_{\vec{k}}\int_{-\infty}^{\infty}\dd b_{\vec{k}}~P[a_{\vec{k}},b_{\vec{k}}] = 1, &&\forall \vec{k}.
\end{align}
Moments of any functional $Q[f(\vec{k})]$ of the random field can be easily computed using the characterisation above:
\begin{align*}
	\mathbb{E}[Q[f(\vec{k})]]=\int\text{D}f(\vec{k}) ~P[f(\vec{k})]~Q[f(\vec{k})]= \int_{-\infty}^{\infty}\dd a_{\vec{k}}\int_{-\infty}^{\infty}\dd b_{\vec{k}}~\frac{Q[a_{\vec{k}},b_{\vec{k}}]}{\pi\sigma^2_{\vec{k}}}e^{-\frac{a_{\vec{k}}^2+b_{\vec{k}}^2}{\sigma^2_{\vec{k}}}}.
\end{align*}
This can then be Fourier transformed to real space. As a simple example, lets compute a two-point function of the random field,
\begin{align*}
	\mathbb{E}[f(\vec{x})f(\vec{y})]&=\int\frac{\dd[d] k}{(2\pi)^d}\int\frac{\dd[d] q}{(2\pi)^d} e^{i\vec{k}\cdot\vec{x}}e^{i\vec{q}\cdot\vec{y}} \mathbb{E}[f(\vec{k})f(\vec{q})]\\
	&=\int\frac{\dd[d] k}{(2\pi)^d}\int\frac{\dd[d] q}{(2\pi)^d} e^{i\vec{k}\cdot\vec{x}}e^{i\vec{q}\cdot\vec{y}} \sigma^2_{\vec{k}}\delta(\vec{k}+\vec{q})\\
	&=\int\frac{\dd[d] k}{(2\pi)^d}e^{i\vec{k}\cdot(\vec{x}-\vec{y})}\frac{\sigma^2_{\vec{k}}}{(2\pi)^d},
\end{align*}
where $(2\pi)^{-d}\sigma^2_{\vec{k}}$ is commonly known as the \emph{power spectrum} of the random field. For the simple case where the power spectrum is a constant $\sigma^2_{\vec{k}}=(2\pi)^{d}\bar{V}^2$, we have $\mathbb{E}[f(\vec{x})f(\vec{y})]=\bar{V}^2\delta(x-y)$ which means the distribution of $f(\vec{x})$ in real space is also Gaussian. Conversely, considering a non-trivial power spectrum lead to non-trivial correlations for the random field. Therefore the spectral representation of a random field is a convenient way of generating non-gaussian distributions while still working with Gaussian objects in Fourier space.

\subsection{Cutoffs}
However this construction is not very useful for actual applications. For instance, note that as $\vec{x}$ approaches $\vec{y}$ we get an ultra-violet divergence. This for instance can be very inconvenient in the case we are treating, since in our equations we have a lot of terms that go as 'disorder squared' at the same point. To remediate this problem, we will resolve this long wavelength divergence by introducing an ultra-violet (UV) cutoff $\lambda$ and integrate only over modes $|\vec{k}|<\lambda$. A pictorial way to interpret this cutoff is to say that $\lambda$ introduce a length scale $a=\pi/\lambda$ that corresponds to an underlying lattice. Modes with frequencies below this scale are then ignored. This would
	imply for instance that $\mathbb{E}[f(\vec{x})^2] = \bar{V}^2/a$. As we take the lattice spacing $a\to 0$ we recover the expected UV divergence.
	 
While UV divergences are a consequence of the way we introduce disorder, there can be IR divergences that are emergent in the problem, and indicate a change of behaviour in the system. Or in terms of the renormalisation group: the system flows towards a disordered fixed point. To resolve these divergences one usually introduce a box of size $L$, and in the end of calculation one aims to study how the system behaves as $L$ is increased. IR singularities in the thermodynamic limit $L\to\infty$ indicate a flow towards a new phase.

\subsection{Discrete}
\label{discrete}
Although the continuum implementation simplifies analytical calculations, numerically one needs a discretisation that takes into account the aforementioned observations. From now on we restrict ourselves to the case of interest, namely $d=1$. Note that according to Eq. \eqref{distribution} the norm of $|f(\vec{k})|$ is drawn from a Gaussian distribution, while the phase is drawn from the uniform distribution on $[0,2\pi]$. This remark leads to a useful discrete representation of the continuous spectral decomposition.
We start by discretising uniformly our box of size $L$ in $N$ intervals of size $a$, i.e. $L= Na$. This implies the quantisation of the modes $k_n = n\Delta k$ with $\Delta k = \frac{\pi}{N a} = \frac{\lambda}{N}$. The thermodynamic limit then becomes $N\to\infty$. The spectral decomposition Eq. \eqref{spectral} becomes,
\begin{align}
\label{dspectral}
	f(x) = \sum\limits_{n=-N}^{N}f_n e^{ik_n x} = \sum\limits_{n=1}^{N}A_n \cos(k_n x+\gamma_n),
\end{align}
where $A_n\in\mathbb{R}_{>0}$ corresponds to the amplitude of the Fourier modes and $\gamma_n\in[0,2\pi]$ the phase. As we remarked above, in this representation each $A_n$ is a random variable taking values in a Gaussian distribution, while each $\gamma_n$ is a random variable taking values uniformly in $[0,2\pi)$. However a useful simplification is to make $A_n$ deterministic and only keep the phases random. One can then check that taking $A_n = \bar{V}\sqrt{\sigma(k_n)\Delta k}$ and averaging uniformly in $[0,2\pi)$,
\begin{align}
\label{eq:appA:average}
\mathbb{E}[\cdots] = \lim\limits_{N\to\infty}\int_{0}^{2\pi}\prod\limits_{n=1}^{N}\frac{\dd \gamma_n}{2\pi}(\cdots),
\end{align}
\noindent imply that for $\sigma(k_n)=1$ in the thermodynamic limit $\mathbb{E}[f(x)f(y)] = \bar{V}^2\delta(x-y)$. In other words, we reproduce the Gaussian behaviour with a simpler setup in which only the phases fluctuate. Similarly, we can obtain non-Gaussianity by choosing a non-trivial function $\sigma(k_n)$.

%%%%%%%%%%%%%%%%%%%%%%%%%%%%%%%%%%%%%%%%%%%%%%%%%%%%%%%%%%
\section{Holographic Renormalisation}
%%%%%%%%%%%%%%%%%%%%%%%%%%%%%%%%%%%%%%%%%%%%%%%%%%%%%%%%%%
\label{app:B}

In this Appendix we discuss the details of holographic renormalisation in our inhomogeneous geometry. Let $\mathcal{M}$ be the underlying manifold defined by the geometry in Eq.\eqref{eq:bulk_metric}. The action for the probe scalar is given by
\begin{align}
S[\psi]&=  \frac{1}{2}\int_{\mathcal{M}}\dd[3] x \left(\dd\psi\wedge \star\dd\psi +m^2 \psi^2\right)\notag\\
&=-\frac{1}{2}\int_{\mathcal{M}}\dd[3] x~ \psi\left(\dd\star\dd\psi-m^2\psi\right)+\frac{1}{2}\int_{\partial\mathcal{M}}\dd[2] x~\psi \star \dd\psi.
\end{align}
Thus on-shell we have
\begin{align}
\label{eq:onshell}
S_{\text{on-shell}}[\psi] = \frac{1}{2}	\int_{\partial\mathcal{M}}\dd[2] x~\sqrt{-\gamma}~\psi n^a \partial_a \psi,
\end{align}
\noindent where $n$ is the normal unit vector pointing outwards of the boundary $\partial\mathcal{M}$ and $\gamma$ is the respective induced metric. Note that strictly at the conformal boundary $\rho=0$ Eq.\eqref{eq:onshell} is divergent. To regularise this divergence, we evaluate the on-shell action at a slice $\rho=\lambda \ll 1$ which we later take to zero. The normal unit vector pointing outward the fixed $\rho=\lambda$ surface is then given by $n = -2\rho~ \partial_{\rho}|_{\rho=\lambda}$, while the induced metric is 
\begin{align}
\dd s^2=\lambda^{-1}g_{(0)}(x)\dd x^\mu \dd x^{\nu}=\lambda^{-1}\left(-\dd t^2 + \dd x^2+2g_{tx}(x)\dd t\dd x\right)	 
\end{align}
\noindent and thus $\sqrt{-\gamma} = \lambda^{-1}\sqrt{-g_{(0)}}$ with $\sqrt{-g^{(0)}}=\sqrt{1+g_{tx}^2}$. Note that $g^{(0)}$ is interpreted holographically as the metric where the dual boundary field theory lives. Inserting in the on-shell action,
\begin{align}
S_{\text{on-shell}}[\psi] = -\int\dd t \dd x ~\sqrt{-g_{(0)}}~\psi \partial_\rho \psi|_{\rho=\lambda}.
\end{align}
As with the pure AdS case, the above on-shell action needs to be renormalised. Solutions to Eq. \eqref{eq:scalar_eom} satisfy the following near-boundary expansion
\begin{align}
\psi(\rho,x)=\rho^{\frac{1-\nu}{2}} s(x)+\rho^{\frac{1+\nu}{2}}A(x)+\dots,
\end{align}
\noindent where all higher order terms have positive exponents. Thus we have
\begin{align}
S_{\text{on-shell}}[\psi] &= -\int\dd t \dd x ~\sqrt{-g_{(0)}}\Big(\lambda^{\frac{1-\nu}{2}}s(x)+\lambda^{\frac{1+\nu}{2}}A(x)+\dots\Big)\times\notag\\
&\hspace{9em}\times\Big(\frac{1-\nu}{2}\lambda^{-\frac{1+\nu}{2}}s(x)+\frac{1+\nu}{2}\lambda^{\frac{\nu-1}{2}}A(x)+\dots\Big),\notag\\
&=-\int\dd t \dd x ~\sqrt{-g_{(0)}}\left(\frac{1-\nu}{2}\lambda^{-\nu}s(x)^2+s(x)A(x)+\dots\right).
\end{align}
Note that the highest diverging is in the term $\propto s^2$. To remediate this particular divergence, we add the following boundary counter-term to the action
\begin{align}
	S_{\text{ct}}^{(0)}&=\frac{1-\nu}{2}\int_{\partial\mathcal{M}}\dd[2] x\sqrt{-\gamma}~\psi^2|_{\rho=\lambda} =\frac{1-\nu}{2}\int\dd t\dd x\sqrt{-g_{(0)}}~\left(\lambda^{-\nu} s(x)^2+2 s(x)A(x)+\dots\right).
\end{align}
In most of the manuscript we work with $\nu = 1/2$, for which other divergences are not present, and therefore the only counter term required is the above. However, for larger values of $\nu$ there is a finite tower of higher divergences between the leading term $\propto \int s(x)^2$ and the term $\propto \int s(x) A(x)$ which we have omitted in the ellipsis, and which should also be taken into account in order to obtain a finite one-point function. For instance, the next order divergence would be of order $O\left(\lambda^{-(\nu-1)}\right)$, which can be remediated by adding a derivative term $S^{(1)}_{\text{ct}}=\frac{1}{2\nu-2}\int_{\partial\mathcal{M}}\dd^2\sqrt{-\gamma}\phi~\Delta_{\gamma}\phi |_{\rho=\lambda}$. Higher order terms can be treated in the same way, adding higher derivative terms $S^{(k)}_{\text{ct}}$ accordingly (for a complete discussion, see \cite{Skenderis2002}). However note these derivative terms do not contribute to the coefficient of the one-point function $\propto \int s(x)A(x)$. Accounting for all the divergences, the renormalised action reads
\begin{align}
\label{action:renom}
S_{\text{ren}} = \lim\limits_{\lambda\to 0}\left(S_{\text{on-shell}}+S_{\text{ct}} \right)= -\frac{1}{2} \int\dd t \dd x ~\sqrt{-g_{(0)}}~2\nu~s(x)A(x).
\end{align}
The expression above makes clear that the leading coefficient $s(x)$ in the expansion of $\psi$ acts as a source for a dual operator $\langle\mathcal{O}(x)\rangle = 2\nu ~A(x)$. The expectation value of the dual is then simply given by
\begin{align}
\langle\mathcal{O}(x)\rangle = \frac{1}{\sqrt{-g_{(0)}}}\frac{\delta\left(-S_{\text{ren}}\right)}{\delta s(x)} 	=2\nu~ A(x) %=2\nu~ \lim\limits_{\rho\to 0}\rho^{\frac{1+\nu}{2}}\psi(\rho,x).
\end{align}

%%%%%%%%%%%%%%%%%%%%%%%%%%%%%%%%%%%%%%%%%%%%%%%%%%%%%%%%%%%%%%%%%%%%%%%%%%
\section{Boundary-to-bulk propagator and boundary two-point function}
%%%%%%%%%%%%%%%%%%%%%%%%%%%%%%%%%%%%%%%%%%%%%%%%%%%%%%%%%%%%%%%%%%%%%%%%%%
\label{app:C}

Consider a probe scalar field $\psi$ living in an asymptotically AdS$_{d+1}$ geometry with metric tensor $g$. As previously discussed, the equation of motion for the scalar is given by $\left(\Delta_g-m^2\right)\psi=0$. We define the \emph{bulk-to-bulk propagator} $G(x,y)$ as the Green's function for this equation. In other words, it is the solution of 
\begin{align}
\label{bulktobulk}
	(\Delta_g - m^2)G(x,y) = \frac{i}{\sqrt{-g}}\delta^{d+1}(x-y).
\end{align}
Note that if we couple the scalar field with a source by adding a term $(\Delta_g - m^2)\psi = J$, then knowing bulk-to-bulk propagator we can build a solution 
\begin{align}
\psi(x) = \int\dd[d+1] y\sqrt{-g}~ G(x,y)J(y).
\end{align}
In a neighbourhood of the boundary, by the Fefferman-Graham theorem we can write the asymptotically AdS metric in a coordinate chart $x^a = (\rho, x^{\mu})$ as
\begin{align}
\dd s^2 &= \frac{\dd\rho^2}{4\rho^2}+\frac{1}{\rho}g_{\mu\nu}(\rho,x^{\mu}) \dd x^{\mu}\dd x^{\nu}, \notag\\
g(\rho,x^{\mu})&\underset{\rho=0}{\sim} g_{(0)}(x^{\mu})+\rho g_{(1)}(x^{\mu})+O(\rho^2),
\end{align}
\noindent where $g_{(0)}(x^{\mu})$ defines the metric at the conformal boundary located at $\rho=0$. At a slice close to the boundary $\rho=\lambda\ll 1$, we have 
\begin{align}
\sqrt{-g} = \frac{\sqrt{-g_{(0)}}}{2\lambda^{1+d/2}}.
\end{align}
And therefore evaluating Eq.\eqref{bulktobulk} at $\rho=0$ for one of the arguments make the right-hand side zero. This defines the so called \emph{Boundary-to-bulk propagator}
\begin{align}
\label{boundarytobulk}
	(\Delta_g - m^2)K(\rho;x^{\mu},y^{\mu}) = 0
\end{align}
\noindent which depends only on one radial variable. As we will see next, it propagates solutions from the boundary to the bulk. Recall that according to the holographic dictionary solutions of the bulk equations of motion define a dual source at the boundary according to
\begin{align}
\lim\limits_{\rho\to 0}\rho^{-\Delta_{-}/2}\psi(\rho,x^{\mu})	= s(x^{\mu}),
\end{align}
\noindent where $\Delta_{\pm} = \frac{d}{2}\pm \nu$ with $\nu = \sqrt{\frac{d^2}{4}+m^2}$. Thus by imposing boundary conditions
\begin{align}
\lim\limits_{\rho\to 0}\rho^{-\Delta_{-}/2}	K(\rho;x^\mu,y^\mu) = \delta^{d}(x^\mu-y^\mu),
\end{align}
\noindent we find that the $K$ satisfies
\begin{align}
\psi(\rho,x^{\mu}) = \int\dd[d] y~K(\rho;x^\mu,y^\mu)s(y^\mu).
\end{align}
This justifies the terminology boundary-to-bulk propagator, since it propagates the source $s(x^\mu)$ living in the boundary into a scalar field satisfying the Bulk equations of motion.

From this relation it is also possible to see that the bulk-to-bulk and boundary-to-bulk propagators are related to the boundary tree-level boundary two-point function. Since we have
\begin{align}
\langle \mathcal{O}(x^\mu)\rangle = 	(2\nu) \lim\limits_{\rho\to 0}\rho^{-\Delta_{+}/2}\psi(\rho,x^\mu)
\end{align}
\noindent we see that if we define the boundary tree-level two-point function 
\begin{align}
G_{(0)}(x^\mu,y^\mu)=\langle\mathcal{O}(x^\mu)\mathcal{O}(y^\mu)\rangle = (2\nu)\lim\limits_{\rho\to 0}\rho^{-\Delta_{+}/2}K(\rho;x^\mu,y^\mu)
\end{align}
\noindent it satisfies the usual linear-response relation
\begin{align}
\langle \mathcal{O}(x^\mu)\rangle = \int \dd[d] y~ G_{(0)}(x^\mu,y^\mu)s(y^\mu)	
\end{align}
\noindent between the source and the expected value. 

\end{document}